\documentclass[11pt,a4paper]{article}

\usepackage[top=20mm,bottom=20mm,left=30mm,right=20mm]{geometry}

\usepackage[dvips]{graphicx}
\graphicspath{{eps/}}

\usepackage[cp1251]{inputenc}
\usepackage[T2A]{fontenc}
\usepackage[russian]{babel}

\usepackage{amsmath}
\usepackage{amssymb}
\usepackage{amsxtra}
\usepackage{amsthm}
\usepackage{latexsym}
\usepackage{array}
\usepackage{multirow}
\usepackage{hhline}
\usepackage{color}
\usepackage{longtable}
\numberwithin{equation}{section}
\theoremstyle{plain}
\newtheorem{theorem}{Теорема}
\newtheorem{lemma}{Лемма}
\newtheorem{propos}{Предложение}
\theoremstyle{remark}
\newtheorem{remark}{Замечание}
\def\mcm{\mathcal{N}}
\def\zu{\zeta}
\def\wu{W}
\def\dd{\delta}
\def\va{\alpha}
\def\vvp{p}
\def\vo{\omega}
\def\vO{\Omega}
\def\bm{\mathbf{M}}
\def\bp{\mathbf{p}}
\def\bal{{\boldsymbol\alpha}}
\def\bom{{\boldsymbol\omega}}
\def\mg{\mathfrak{g}}
\def\po{I}
\def\mP{\mathcal{P}}
\def\mPel{{\mP^4_{a,\ell}}}
\def\roo{\varrho}
\def\ld{\lambda}
\def\ds{\displaystyle}
\def\gs{\geqslant}

\def\iso{Q_{\ell,h}^3}
\def\sgrad{\mathop{\rm sgrad}\nolimits}
\def\mtA{\mathbb{A}}
\def\mtB{\mathbb{B}}
\def\mtC{\mathbb{C}}
\def\mtD{\mathbb{D}}
\def\mtE{\mathbb{E}}
\def\mtF{\mathbb{F}}
\def\mtG{\mathbb{G}}
\def\bR{\mathbb{R}}
\def\eps{\varepsilon}
\def\sgn{\mathop{\rm sgn}\nolimits}
\def\ri{\mathrm{i}\,}
\def\vk{\varkappa}

\newcommand{\fts}[1] {{\small #1}}
\newcommand{\fns}[1]{{\footnotesize #1}}

\usepackage{hyperref}
\begin{document}

\title{Фазовая топология волчка Ковалевской\,--\,Соколова}

\author{П.\,Е.~Рябов, А.\,Ю.~Савушкин}

\date{}

\maketitle

\begin{abstract}
Исследуется фазовая топология интегрируемой гамильтоновой системы на $e(3)$, найденной В.В.Соколовым (2001) и обобщающей случай Ковалевской. Обобщение состоит в том, что к однородному потенциальному силовому полю добавлены гироскопические силы, зависящие от конфигурационных переменных. Классифицированы относительные равновесия, вычислен их тип, определен характер устойчивости. Установлены виды диаграмм Смейла и дана классификация изоэнергетических многообразий приведенных систем с двумя степенями свободы. Множество критических точек полного отображения момента представлено в виде объединения критических подсистем, каждая из которых при фиксированных физических параметрах является однопараметрическим семейством почти гамильтоновых систем с одной степенью свободы. Для всех критических точек явно вычислены показатели, определяющие их тип. Выписаны уравнения поверхностей, несущих бифуркационную диаграмму отображения момента. Приведены примеры изоэнергетических диаграмм с полным описанием соответствующей грубой топологии (регулярных торов Лиувилля и их бифуркаций).

\vspace{3mm}

Ключевые слова: интегрируемые гамильтоновы системы, относительные равновесия, изоэнергетические поверхности, критические подсистемы, бифуркационные диаграммы, грубая топология

\end{abstract}

{

\parindent=0mm

УДК 517.938.5+531.38


------------------------------------------------------------

Получено 26 апреля 2015 гoда

После доработки 19 мая 2015 года

\vspace{5mm}

{\Large Нелинейная динамика. 2015. Т. 11. № 2. С. 287–317.\\ Полнотекстовая версия в свободном доступе \href{http://nd.ics.org.ru}{\path{http://nd.ics.org.ru}}
}

------------------------------------------------------------

Работа выполнена при поддержке гранта РФФИ № 14-01-00119 и совместного гранта РФФИ и АВО № 15-41-02049.

------------------------------------------------------------

Рябов Павел Евгеньевич

peryabov@fa.ru

Финансовый университет при Правительстве Российской Федерации

125993, Россия, г. Москва, Ленинградский проспект, д. 49

------------------------------------------------------------

Савушкин Александр Юрьевич

a\_savushkin@inbox.ru

Российская академия народного хозяйства и государственной службы

400131, Россия, г. Волгоград, ул. Гагарина, д. 8

------------------------------------------------------------

}

\clearpage

{
\linespread{0.3}

\tableofcontents

}

\section{Исходные соотношения и постановка задачи}\label{sec1}
Коалгебра $\mg_0=e(3)^*$ реализуется как $\bR^6(\bm,\bal)$ со скобкой Пуассона
\begin{equation}\label{eq1_1}
  \{M_i,M_j\}=\eps_{ijk}M_k, \quad \{M_i,\va_j\}=\eps_{ijk}\va_k, \quad \{\va_i,\va_j\}=0.
\end{equation}
Соответствующие уравнения Гамильтона
\begin{equation}\label{eq1_2}
  \dot x =\{H,x\},
\end{equation}
в которых $H$ -- заданная функция от $\bm,\bal$ (гамильтониан), записанные в переменных $M_i,\va_j$, называются уравнениями Эйлера\,--\,Пу\-ас\-сона. Заметим, что именно такой порядок аргументов в скобке \eqref{eq1_2} необходим для совпадения знаков с классическими аналогами.

Скобка \eqref{eq1_1} обладает двумя функциями Казимира
\begin{equation}\label{eq1_4}
  L=\frac{1}{2} \bm \boldsymbol{\cdot} \bal, \qquad \Gamma= \bal^2.
\end{equation}
Точкой обозначено скалярное произведение в $\bR^3$, коэффициент в $L$ введен по традиции, сложившейся в задачах динамики твердого тела с конфигурацией типа Ковалевской.

На совместном уровне
\begin{equation*}
  \mPel = \{L=\ell,\Gamma=a^2\}
\end{equation*}
скобка \eqref{eq1_1} невырождена и ограничение системы \eqref{eq1_2} становится гамильтоновой системой с двумя степенями свободы. Иногда нам будет удобно считать фазовым пространством системы \eqref{eq1_2} пятимерное многообразие $\mP^5=\bR^3(\bm){\times}S^2(\bal)$, заданное одним уравнением
\begin{equation*}
  \bal^2 = a^2 \qquad (a>0),
\end{equation*}
и говорить об однопараметрическом (параметр $\ell\in\bR$) семействе систем на $\mPel$. Последнее соотношение в механике называют геометрическим интегралом, определенную им сферу -- сферой Пуассона. Функцию $L$ и порожденное ей соотношение $L=\ell$ называется интегралом площадей.

В работе \cite{So2002} для системы \eqref{eq1_2} найдено одновременное обобщение интегрируемого гиростата Ковалевской \cite{Yeh} и интегрируемой системы Соколова для уравнений Кирхгофа \cite{So2001}. Этот случай естественно называть гиростатом Ковалевской\,--\,Соколова. В настоящей статье мы рассмотрим задачу с гамильтонианом
\begin{equation}\label{eq1_3}
  H=\frac{1}{4}(M_1^2+M_2^2+2M_3^2)+\eps_1(\va_3 M_2-\va_2 M_3)-\eps_0\va_1.
\end{equation}
По сравнению с общим гамильтонианом работы \cite{So2002} здесь отсутствует линейное слагаемое вида $\lambda M_3$, которое характерно для задач о движении гиростата. Поэтому соответствующую
систему будем называть интегрируемым волчком Ковалевской\,--\,Соколова. Задача характеризуется тремя параметрами, которые можно назвать физическими. Это -- $a,\eps_0,\eps_1$. Заметим, что в случае общего положения ($a\eps_0\eps_1 \ne 0$) эта тройка избыточна. Введением подходящих единиц измерения можно два параметра из трех сделать равными единице (кроме пары $\eps_0,\eps_1$, в которой отношение $\eps_1/\eps_0$ является существенным). Однако мы пока сохраним все три параметра, что дает возможность предельных переходов $\eps_1 \to 0$ (классический случай Ковалевской), $\eps_0 \to 0$ (случай Соколова для уравнений Кирхгофа \cite{So2001} и, при введении дополнительного параметра в скобку Пуассона, случай Борисова\,--\,Мамаева\,--\,Соколова на $so(4)$ \cite{BorMamSok2001}) с сохранением произвольного $a>0$. В связи с наличием таких переходов гамильтониан \eqref{eq1_3} иногда называют деформацией случая Ковалевской (см., например, работу \cite{Versh}, где с точки зрения бигамильтоновости обсуждаются различные обобщения случая Ковалевской). Отметим также, что очевидными комбинациями отражений в пространстве $\bR^6$ и инверсии времени можно добиться выполнения неравенств
\begin{equation*}
  \eps_0 > 0, \qquad \eps_1 >0.
\end{equation*}
Так, поворот подвижной системы отсчета на $\pi$ вокруг третьей оси меняет оба знака $\eps_0,\eps_1$, а замена $(M_1,\va_2,\va_3,t) \to (-M_1,-\va_2,-\va_3,-t)$ равносильна замене знака только у $\eps_1$.

Первый интеграл, найденный в \cite{So2002}, дополнительный к $\Gamma,L,H$ и обеспечивающий интегрируемость системы \eqref{eq1_2} (соответственно, лиувиллеву полную интегрируемость семейства гамильтоновых систем на $\mPel$) можно записать в виде
\begin{equation*}
\begin{array}{lll}
K&=& \ds     \left[ \frac{1}{4}(M_1^2-M_2^2)+\eps_1 (\va_2 M_3-\va_3 M_2)-\eps_1^2(\va_1^2+\va_2^2+\va_3^2)+\eps_0\va_1\right]^2+
 \\[3mm]
&+& \ds \left[ \frac{1}{2}M_1 M_2+\eps_1(\va_3 M_1-\va_1 M_3)+\eps_0 \va_2\right]^2.
\end{array}
\end{equation*}

В механике большое внимание уделяется исследованию особых движений механических систем (в том числе и интегрируемых), их аналитическому описанию и изучению характера устойчивости.
Количество классических работ по этой тематике весьма велико. В последнее время вопросы об устойчивости таких движений связываются с топологией соответствующих интегрируемых систем, отображением момента и бифуркационными диаграммами, отражающими все особенности слоений фазового пространства (см., например, работы \cite{BolBorMam1, BorMam3,BolBorMam2}). В динамике твердого тела особое место занимает класс движений, называемых равномерными вращениями, в которых вектор угловой скорости тела постоянен в подвижной и неподвижной системах отсчета. С точки зрения системы уравнений Эйлера\,--\,Пуассона эти движения являются неподвижными точками, поэтому они также называются относительными равновесиями. Устойчивость относительных равновесий в значительной мере определяется собственными числами матрицы правой части линеаризованных уравнений для системы \eqref{eq1_2}. В интегрируемой системе эти собственные числа определяют так называемый тип критической точки \cite{BolFom}, соответствующей относительному равновесию.

С другой стороны, в топологическом анализе интегрируемой системы строятся топологические инварианты (меченые молекулы Фоменко\,--\,Цишанга \cite{FoZi1990}) на изоэнергетических многообразиях $\iso=\{H=h\} \cap \mPel$ систем на $\mPel$. Очевидно, эти многообразия зависят также и от $a$, но эту зависимость явно не пишем. На соответствующие метки оказывает влияние сама топология многообразий $\iso$. Поэтому важным этапом топологического анализа является классификация изоэнергетических многообразий. Соответствующий математический аппарат разработан Смейлом \cite{Smale}. Оказывается, что перестройки топологического типа $\iso$ происходят при пересечении значений параметров, отвечающим относительным равновесиям, а вид перестройки определен индексом Морса ограничения функции $H$ на $\mPel$ в точках относительных равновесий. При фиксированных физических параметрах, множество разделяющих значений для топологического типа $\iso$ в плоскости $(\ell,h)$ называют диаграммой Смейла.

Ниже получено полное аналитическое описание всех относительных равновесий волчка Ковалевской\,--\,Со\-ко\-ло\-ва, вычислены типы относительных равновесий и установлен характер их устойчивости. Перечислены виды диаграмм Смейла, вычислены индексы Морса приведенной энергии и установлена топология изоэнергетических уровней.

В целом множество критических точек отображения момента рассматриваемой системы получено как объединение четырех критических подсистем. Поскольку, как отмечалось, фазовое пространство системы в целом расслоено на симплектические листы -- четырехмерные фазовые пространства приведенных систем, критические подсистемы также представляют собой объединения однопараметрических семейств ($a$ фиксировано, $\ell$ произвольно) двумерных фазовых пространств индуцированных почти гамильтоновых систем с одной степенью свободы. Каждое из этих пространств имеет подмножество коразмерности 1, на котором вырождается форма, индуцированная исходной симплектической структурой. Получены условия вырождения в виде равенства нулю некоторого частного интеграла. Явно вычислен тип критических точек ранга 1. В частности, в терминах констант интегралов выписаны уравнения множеств вырожденных критических точек.

Полученная информация, вместе с результатами \cite{Ry2013,KhRCD2014} по фазовой топологии обобщенного двухполевого гиростата Соколова\,--\,Цыганова \cite{SoTs2002}, дает возможность построить бифуркационные диаграммы отображения момента, оснащенные обозначениями бифуркаций и указанием количества регулярных торов Лиувилля, что и определяет грубую топологию системы. Знание грубой топологии в целом позволяет завершить и полное описание относительных равновесий указанием топологической структуры их насыщенной четырехмерной окрестности в фазовом пространстве приведенной системы в тех случаях, когда по аналитически найденному типу точки эта структура не устанавливается однозначно.

\section{Множество относительных равновесий}\label{sec2}
Для дальнейшего удобно использовать наряду с моментами и компоненты угловой скорости $\displaystyle \bom={\partial H}/{\partial \bm}$:
\begin{equation*}
  \vo_1=\frac{M_1}{2},\qquad   \vo_2=\frac{M_2}{2}+\eps_1\va_3,\qquad   \vo_3=M_3-\eps_1 \va_2.
\end{equation*}
Первые интегралы примут вид
\begin{equation}\label{eq2_1}
\begin{array}{lll}
  L&=&\ds \va_1 \vo_1+ \va_2 \vo_2+\frac{1}{2} \va_3 \vo_3- \frac{1}{2}\eps_1\va_2\va_3 = \ell,\\[3mm]
  H&=&\ds \vo_1^2+\vo_2^2+\frac{1}{2}\vo_3^2-\frac{1}{2}\eps_1^2(\va_2^2+2\va_3^2)-\eps_0\va_1 =h,\\[3mm]
  K&=&\ds \left[\vo_1^2-\vo_2^2+\eps_1 \va_2 \vo_3+(\eps_0-\eps_1^2\va_1)\va_1\right]^2+\\[2mm]
  &+&\ds \left[2\vo_1\vo_2 -\eps_1 \va_1 \vo_3+(\eps_0-\eps_1^2\va_1)\va_2\right]^2=k.
\end{array}
\end{equation}

Гамильтоново поле, порожденное произвольной функцией $F$ с помощью заданной скобки Пуассона, обозначают через $\sgrad F$. Поэтому векторное поле, отвечающее системе \eqref{eq1_2}, есть $\sgrad H$. В координатах $\vo_i,\va_j$ получаем
\begin{equation*}
\begin{array}{lll}
  \sgrad H &=&\ds \Bigl( \frac{1}{2}(\vo_2-\eps_1\vo_3)(\vo_3+\eps_1 \vo_2), -\frac{1}{2}\left[(\eps_0-2\eps_1^2\va_1)\va_3+(\vo_3+\eps_1 \va_2)\vo_1\right],   \Bigr. \\
  &{}& \ds \quad \Bigl. (\eps_0-\eps_1^2\va_1)\va_2+\eps_1 \va_3\vo_1, \va_2\vo_3-\va_3\vo_2, \va_3\vo_1-\va_1\vo_3,\va_1\vo_2-\va_2\vo_1 \Bigr).
\end{array}
\end{equation*}

Неподвижные точки системы Эйлера\,--\,Пуассона определяются из условия ${\sgrad H=0}$. В частности, существует скалярная константа $\vO$, такая, что
\begin{equation}\label{eq2_2}
  \bom = \vO \bal.
\end{equation}
Оставшиеся условия относительного равновесия дают
\begin{eqnarray}
& (\vO \va_2-\eps_1 \va_3)(\vO \va_3+\eps_1 \va_2)=0,  \label{eq2_3}\\
& \eps_1\vO \va_1 \va_2+[\eps_0+(\vO^2-2\eps_1^2)\va_1]\va_3=0,\quad
(\eps_0 -\eps_1^2\va_1)\va_2+\eps_1 \vO\va_1 \va_3 =0. \label{eq2_4}
\end{eqnarray}

Для дальнейшего примем следующую точку зрения. Ненулевые параметры $\eps_0,a$ в совокупности характеризуют взаимодействие волчка с потенциальным силовым полем. Будем считать их выбранными и фиксированными. Классифицирующим различные системы будем считать параметр $\eps_1$, характеризующий гироскопические силы и обеспечивающий деформацию задачи Ковалевской. Различие систем будет удобно определять по величине $\eps_1^2$. Введем обозначения для значений этой величины, которые будут в разных ситуациях служить разделяющими:
\begin{equation}\label{eq2_5}
\ds   \zeta_1 = \frac{\eps_0}{2a}, \qquad \zeta_2 = \frac{\eps_0}{a}, \qquad \zeta_3 = (5+3\sqrt{3})\frac{\eps_0}{a}.
\end{equation}

Опишем относительные равновесия в терминах вектора $\bal$ и величины $\vO$ с учетом равенства \eqref{eq2_2}.

\begin{propos}\label{prop1}
$({\rm i})$ Относительные равновесия волчка Ковалевской\,--\,Соколова образуют следующие семейства:
\begin{equation}\label{eq2_6}
\begin{array}{llll}
\dd_{1,2}: & \ds\va_1 =\pm a, & \ds\va_2=0, & \ds\va_3=0, \\[3mm]
\dd_{3}: & \ds\va_1 =\frac{\eps_0}{\eps_1^2-\vO^2},& \ds\va_2= \frac{\eps_1}{\eps_1^2-\vO^2} R_3(\vO),& \ds\va_3= \frac{\vO}{\eps_1^2-\vO^2} R_3(\vO),\\[3mm]
\dd_4: & \ds\va_1 =\frac{\eps_0}{2\eps_1^2},& \ds\va_2= -\frac{\vO}{2\eps_1^2} R_4(\vO),&
    \ds\va_3= \frac{1}{2\eps_1^2} R_4(\vO),
\end{array}
\end{equation}
где
\begin{equation*}
R_3^2(\vO) = \frac{a^2(\eps_1^2-\vO^2)^2-\eps_0^2}{\eps_1^2+\vO^2}, \qquad
R_4^2(\vO) = \frac{4 \eps_1^2 a^2-\eps_0^2}{\eps_1^2+\vO^2}.
\end{equation*}

$({\rm ii})$ В семействах $\dd_1,\dd_2$ величина $\vO$ произвольна. Семейство $\dd_3$ состоит из двух подсемейств $\dd'_3$ для $\vO^2\in [\eps_1^2+\zeta_2,+\infty)$ и $\dd''_3$ для $\vO^2\in [0,\eps_1^2-\zeta_2]$. Первое подсемейство существует при всех значениях физических параметров, второе -- только при условии $\eps_1^2 > \zeta_2$. Семейство $\dd_4$ существует только при условии
$\eps_1^2 > \zeta_1$, и в нем величина $\vO$ произвольна.

$({\rm iii})$ При фиксированных физических параметрах, таких, что $\eps_1^2 \ne \zeta_1$ и $\eps_1^2 \ne \zeta_2$, допустимому значению $\vO$ отвечает по одной точке семейств $\dd_1,\dd_2$ и по две точки семейств $\dd_3,\dd_4$.
\end{propos}

Отметим, что при равенстве $\eps_1^2 =\zeta_1$ все семейство $\dd_4$ сливается с $\dd_1$, а граничный случай $\eps_1^2 = \zeta_2$ приводит к единственной точке в $\dd''_3$, которая принадлежит $\dd_1$ (значение $\vO=0$).

Для доказательства предложения достаточно заметить, что уравнения \eqref{eq2_4} образуют линейную однородную систему по $\va_2,\va_3$. Нулевое решение дает семейства $\dd_1,\dd_2$, а для ненулевых решений равенство нулю первого или второго сомножителя в \eqref{eq2_3} приводит, соответственно, к семействам $\dd_3,\dd_4$.

Вычислим значения первых интегралов в точках найденных семейств. Получим
\begin{equation*}
\begin{array}{l}
\ds \dd_1: \quad h=-\eps_0 a+a^2\vO^2, \quad \ell= a^2\vO, \quad k=a^2[\eps_0-a(\eps_1^2-\vO^2)]^2, \\[2mm]
\ds \dd_2: \quad h=\phantom{-}\eps_0 a+a^2\vO^2, \quad \ell= a^2\vO, \quad k=a^2[\eps_0+a(\eps_1^2-\vO^2)]^2,\\[2mm]
\ds \dd_3: \quad h=-\frac{\eps_0^2(\eps_1^2-3\vO^2)+a^2(\eps_1^2-\vO^2)^3}{2(\eps_1^2-\vO^2)^2}, \quad
\ell=\frac{\vO[\eps_0^2+a^2(\eps_1^2-\vO^2)^2]}{2(\eps_1^2-\vO^2)^2}, \quad k=0,\\
\ds \dd_4: \quad h=-\frac{\eps_0^2}{4\eps_1^2}-\eps_1^2 a^2+a^2\vO^2, \quad \ell= a^2\vO, \quad k=\frac{1}{16\eps_1^4}(\eps_0^2+4 \eps_1^2 a^2\vO^2)^2.
\end{array}
\end{equation*}

Из предложения~\ref{prop1} и выражений первых интегралов находим количество относительных равновесий в прообразах.
\begin{propos}\label{prop2}
В прообразе точки из пространства констант первых интегралов, отвечающей случаю наличия относительного равновесия, имеется по одному относительному равновесию для семейств $\dd_1$, $\dd_2$ и по два -- для семейств $\dd_3$, $\dd_4$.
\end{propos}

\section{Диаграммы Смейла}\label{sec4}
В работе \cite{Smale} Смейл поставил вопрос о структуре бифуркационных диаграмм энергии--мо\-мен\-та в системах с симметрией и решил его для натуральных механических систем: бифуркационная диаграмма $\Sigma_{LH}$ отображения
\begin{equation*}
  L{\times}H: TM \to \bR^2
\end{equation*}
($M$ -- конфигурационное пространство) состоит из пар $(\ell,h)$, в которых $h$ -- критическое значение так называемого эффективного потенциала -- зависящей от константы $\ell$ интеграла момента $L$ функции $V_\ell$ на конфигурационном пространстве $\tilde M $ профакторизованной системы.
Далее диаграммы $\Sigma_{LH}$ называем диаграммами Смейла.

В работе \cite{KhMtt83} схема Смейла была распространена на механические системы с гироскопическими силами. К таким системам относится задача о движении твердого тела с неподвижной точкой (конфигурационное пространство $M=SO(3)$) с гамильтонианом \eqref{eq1_3}. Профакторизованная система -- это индуцированная система на $\mP^5=\{(\bm,\bal)\}$, так что $\tilde M =S^2(\bal)$ -- это сфера Пуассона. Эффективный потенциал на сфере задан в этом случае формулой
\begin{equation}\label{eq3_1}
  V_\ell (\bal)=-\eps_0 \va_1 -\frac{1}{2}\eps_1^2(\va_2^2+2\va_3^2)+\frac{(2\ell+\eps_1\va_2 \va_3)^2}{2[2(\va_1^2+\va_2^2)+\va_3^2]}.
\end{equation}
Поскольку критические точки отображения $L{\times}H$ -- это в точности относительные равновесия, то критические точки $V_\ell$ описываются предложением~\ref{prop1}: при заданном $\ell$ нужно выбрать $\vO$ в соответствии с \eqref{eq2_2} так, что $L(\vO \bal,\bal)=\ell$, то есть положить
\begin{equation*}
  \vO=\frac{2\ell+\eps_1 \va_2 \va_3}{2(\va_1^2+\va_2^2)+\va_3^2}.
\end{equation*}

Образы семейств относительных равновесий в плоскости $(\ell,h)$ образуют кривые, формирующие диаграммы Смейла. Допуская некоторую вольность, будем эти кривые обозначать так же, как и сами семейства:
\begin{equation}\label{eq3_2}
\begin{array}{rl}
\dd_1: &  \ds h=-\eps_0 a+a^2\vO^2, \quad \ell= a^2\vO,   \\[2mm]
\dd_2: &  \ds h=\phantom{-}\eps_0 a+a^2\vO^2, \quad \ell= a^2\vO, \\[2mm]
\dd'_3, \dd''_3: & \ds h=-\frac{\eps_0^2(\eps_1^2-3\vO^2)+a^2(\eps_1^2-\vO^2)^3}{2(\eps_1^2-\vO^2)^2}, \quad
\ell=\frac{\vO[\eps_0^2+a^2(\eps_1^2-\vO^2)^2]}{2(\eps_1^2-\vO^2)^2}, \\
\dd_4: &  \ds h=-\frac{\eps_0^2}{4\eps_1^2}-\eps_1^2 a^2+a^2\vO^2, \quad \ell= a^2\vO.
\end{array}
\end{equation}

Напомним обозначения \eqref{eq2_5}.
\begin{propos}\label{prop3}
В случае Ковалевской\,--\,Соколова существует четыре вида диаграмм Смейла, устойчивых относительно малых возмущений параметров. Разделяющими значениями параметров служат $\eps_1^2=\zeta_i$ $(i=1,2,3)$.
\end{propos}

Доказательство следует из свойств кривых $\dd_j$ и их взаимных пересечений. Перечислим необходимые факты.

Нетрудно видеть, что $\dd_1, \dd_2$ и $\dd_4$ -- это конгруэнтные параболы
\begin{equation}\label{eq3_3}
  \dd_i: \quad h = c_i+\frac{\ell^2}{a^2}, \quad \ell \in \bR,
\end{equation}
причем
\begin{equation*}
  c_4 = - \frac{\eps_0^2+4\eps_1^4a^2}{4\eps_1^2} < c_1=-\eps_0 a <c_2=\eps_0 a.
\end{equation*}
Первое неравенство выполнено, естественно, в области параметров $\eps_1^2 > \zeta_1$, где кривая $\dd_4$ существует.

Диаграммы Смейла, очевидно, симметричны относительно оси $\ell=0$, поэтому все дальнейшие рассуждения приводим для $\ell \gs 0$, не оговаривая это особо.

\begin{figure}[!ht]
\centering
\includegraphics[width=0.9\textwidth]{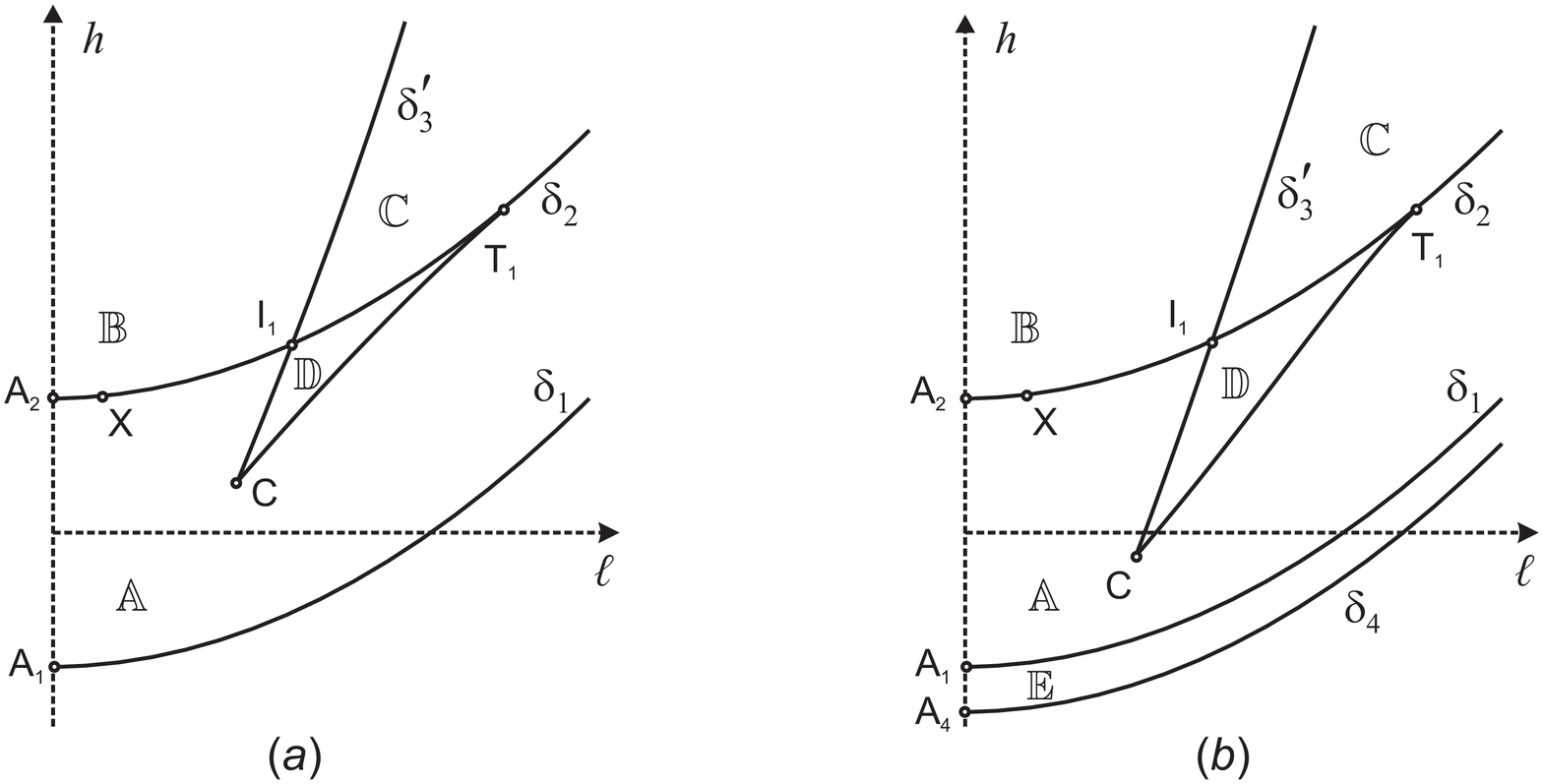} \\[3mm]
\includegraphics[width=0.9\textwidth]{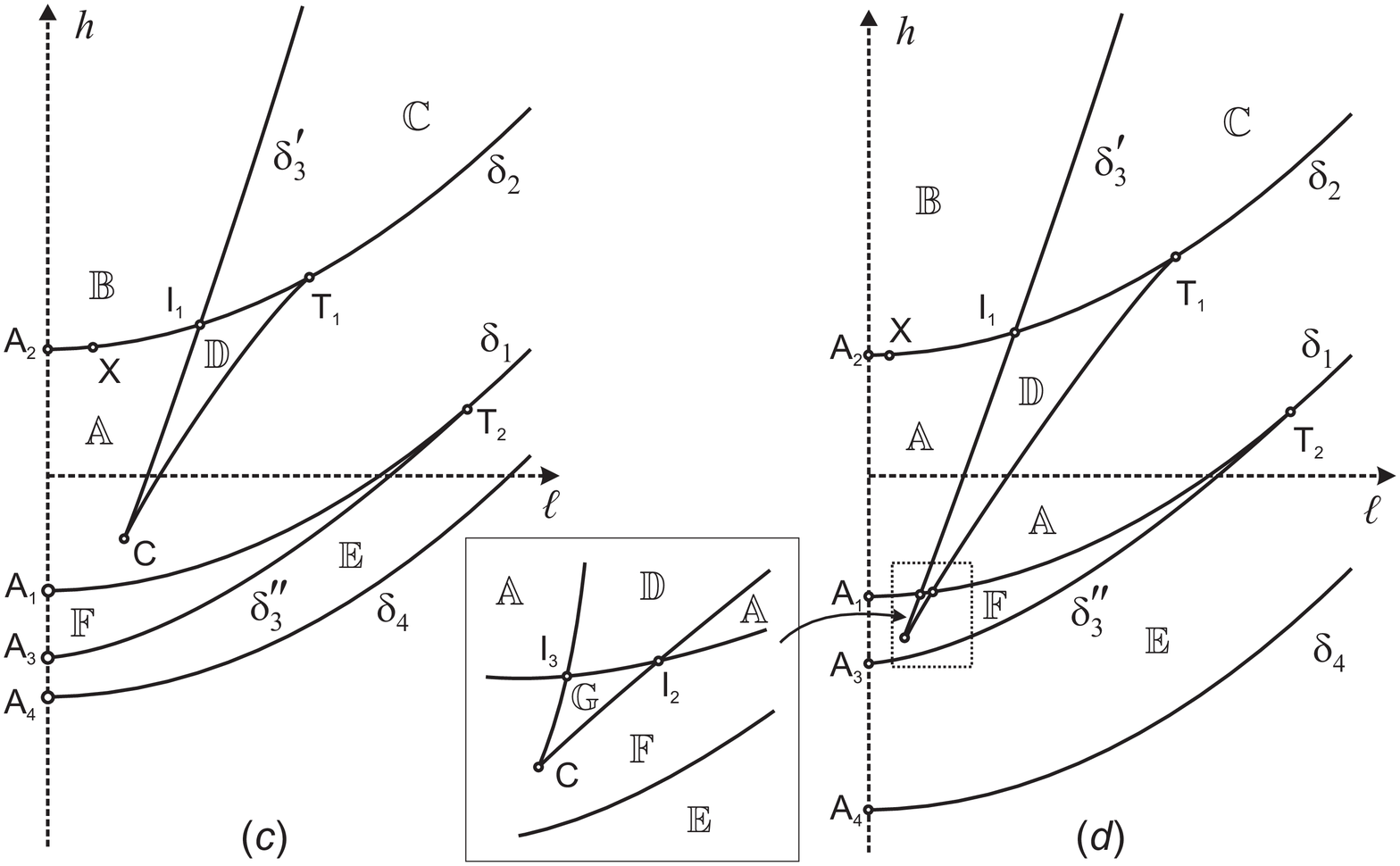}
\caption{Диаграммы Смейла и порожденные области}\label{fig01}
\end{figure}

Кривая $\dd_3$ не имеет простой явной зависимости. Отметим, что ее аналог в классическом случае Ковалевской ($\eps_1=0$) можно записать в виде
\begin{equation*}
  \ell_{\pm} = \frac{1}{27\eps_0^2}\left[h(h^2+9\eps_0^2 a^2) \pm (h^2-3\eps_0^2a^2)^{3/2} \right],
\end{equation*}
так что на $\ell_+$
\begin{equation*}
  h \in \left[\eps_0 a \sqrt{3},2\eps_0 a \right], \qquad \ell^2 \in \left[\frac{4}{3\sqrt{3}}\eps_0a^3,\eps_0 a^3 \right],
\end{equation*}
а на $\ell_-$
\begin{equation*}
  h \in \left[\eps_0 a \sqrt{3},+\infty \right), \qquad \ell^2 \in \left[\frac{4}{3\sqrt{3}}\eps_0a^3,+\infty \right).
\end{equation*}
Сегмент $\ell_+$ соединяет точку возврата $C$ кривой $\dd'_3$
\begin{equation*}
  C: \quad \ell = a \sqrt{\frac{4}{3\sqrt{3}}\eps_0 a}, \quad h =\eps_0 a \sqrt{3}
\end{equation*}
с точкой $T_1$ касания кривых $\dd'_3$ и $\dd_2$
\begin{equation*}
  T_1: \quad \ell = a \sqrt{\eps_0 a}, \quad h =2\eps_0 a.
\end{equation*}
При этом вся кривая $\dd'_3$ находится строго выше кривой $\dd_1$, кривых $\dd''_3,\dd_4$ не существует. Соответствующая диаграмма показана на рис.~\ref{fig01},{\it a}, дополнительные обозначения будут объяснены позже.

Пусть $\eps_1>0$.
\begin{lemma}\label{lem1}
В полуплоскости $\ell>0$ кривая $\dd'_3$ имеет с кривой $\dd_2$ ровно одну точку касания $T_1$ и ровно одну точку пересечения $\po_1$. Кривая $\dd'_3$ имеет единственную точку возврата $C$.
\end{lemma}
\begin{proof}
Записывая условие пересечения кривой $\dd'_3$ с параболой $\dd_2$ в явной форме \eqref{eq3_3} и выполняя подстановку на $\dd'_3$
\begin{equation}\label{eq3_4}
\vO^2=\eps_1^2+\zeta_2+v, \qquad v>0,
\end{equation}
получим уравнение
\begin{equation*}
  v^2 P_1(v)=0,
\end{equation*}
где
\begin{equation*}
P_1(v)=a^3 v^3-(\eps_1^2 a-\eps_0) a^2 v^2-2\eps_0(2\eps_1^2 a+\eps_0) a v -2\eps_0^2(2\eps_1^2 a+\eps_0).
\end{equation*}
Кратный корень $v=0$ отвечает точке касания в начале промежутка изменения $\vO$,
координаты которой вычисляются так:
\begin{equation*}
  T_1: \quad \ell = a \sqrt{(\eps_0+\eps_1^2 a) a}, \quad h =(2\eps_0+\eps_1^2 a) a.
\end{equation*}
Многочлен $P_1(v)$, обладая свойствами $P_1(0)<0$, $P'_1(0)<0$, $P_1(+\infty)=+\infty$, имеет ровно один положительный, и притом простой, корень, отвечающий точке трансверсального пересечения $\dd'_3$ и $\dd_2$, обозначенной через $\po_1$.

Для нахождения точек возврата запишем на $\dd_3$ систему
\begin{equation*}
  \partial _\vO h =0, \qquad \partial _\vO \ell =0.
\end{equation*}
Получим одно уравнение, которое в подстановке \eqref{eq3_4} примет вид $P_2(v)=0$, где
\begin{equation*}
P_2(v) = a^3 v^3+3\eps_0 a^2 v^2-2\eps_0^2(2\eps_1^2a+\eps_0).
\end{equation*}
Очевидно, и этот многочлен на полупрямой $v>0$ имеет единственный и простой корень, определяющий точку возврата $C$.
\end{proof}

Отметим зависимости между $\eps_1$ и $\vO$ (для семейства $\dd'_3$) в точках $\po_1$ и $C$, вытекающие из уравнений $P_1(v)=0$ и $P_2(v)=0$:
\begin{equation}\label{eq3_5}
\left\{ \begin{array}{ll}
\po_1: & \ds \eps_1^2 = \frac{(a v+\eps_0 ) (a^2 v^2-2 \eps_0^2 )}{a (a v+2 \eps_0)^2}\\
C: & \ds \eps_1^2= \frac{a^3 v^3  + 3 \eps_0 a^2 v^2 -2 \eps_0^3}{4 a \eps_0^2}
\end{array} \right. ; \qquad \ds  v=\vO^2-(\eps_1^2+\frac{\eps_0}{a})\gs 0.
\end{equation}
Для точки $\po_1$, принадлежащей также образу семейства $\dd_2$, значение $\vO$ в $\dd_2$ будет другим.

В качественном плане диаграмма Смейла, построенная для малых $\eps_1$, сохранится при увеличении $\eps_1$ до значения $\eps_1^2=\zeta_1$, после которого появляется кривая $\dd_4$ (рис.~\ref{fig01},{\it b}), лежащая строго ниже всех предыдущих кривых.

Дальнейшая эволюция диаграммы Смейла связана с появлением сегмента $\dd''_3$ в момент $\eps_1^2=\zeta_2$. При этом значении $\eps_1$ множество $\dd''_3$ состоит из единственной точки на оси $\ell=0$, принадлежащей $\dd_1$.

\begin{lemma}\label{lem2}
При $\eps_1^2>\zeta_2$  в полуплоскости $\ell>0$ кривая $\dd''_3$ имеет с кривой $\dd_1$ ровно одну точку касания $T_2$ и не имеет  с ней точек пересечения.
\end{lemma}
\begin{proof}
Очевидно, что при таких значениях параметров кривая $\dd''_3$ начинается при $\vO=0$ на оси $\ell=0$ ниже кривой $\dd_1$. Условие пересечения кривой $\dd''_3$ с параболой $\dd_1$, записанной в явной форме \eqref{eq3_3}, в подстановке
\begin{equation}\label{eq3_6}
\vO^2=\eps_1^2-\zeta_2-u, \qquad u>0
\end{equation}
дает уравнение $u^2 P_3(u)=0$, где
\begin{equation*}
P_3(u)=a^3 u^3+(\eps_1^2 a+\eps_0) a^2 u^2+2\eps_0(2\eps_1^2 a-\eps_0) a u +2\eps_0^2(2\eps_1^2 a-\eps_0).
\end{equation*}
Кратный корень $u=0$ отвечает точке касания в конце промежутка изменения $\vO$,
координаты которой вычисляются так:
\begin{equation}\notag
T_2: \quad \ell = a \sqrt{(\eps_1^2 a-\eps_0) a}, \quad h =(\eps_1^2 a-2\eps_0) a.
\end{equation}
Многочлен $P_3(u)$ при $\eps_1^2>\zeta_2$ положительных корней не имеет.
\end{proof}

Легко проверить, что $\dd''_3$ не имеет общих точек с другими кривыми в составе диаграммы Смейла. Отметим еще, что вдоль оси $O\ell$ точка $T_2$ всегда лежит левее точки $T_1$ (рис.~\ref{fig01},{\it c}).

Появление нового (и последнего) разделяющего значения $\eps_1^2=\zeta_3$ связано с возникновением пересечений кривых $\dd'_3$ и $\dd_1$ в тот момент, когда на $\dd_1$ попадает точка возврата $C$ кривой $\dd'_3$.

\begin{lemma}\label{lem3}
В полуплоскости $\ell>0$ кривые $\dd'_3$ и $\dd_1$ не имеют общих точек при $\eps_1^2 < \zeta_3$ и имеют ровно две точки пересечения при $\eps_1^2 > \zeta_3$.
\end{lemma}
\begin{proof}
Условие пересечения $\dd'_3$ с параболой $\dd_1$ в подстановке \eqref{eq3_4} дает, после сокращения на ненулевые сомножители, уравнение
\begin{equation}\label{eq3_7}
a^3 v^3-(\eps_1^2 a -5 \eps_0) a^2 v^2+6 \eps_0^2 a v+2\eps_0^3=0.
\end{equation}
Из него выразим $\eps_1^2$ как функцию $v$. При $v>0$ эта функция имеет единственный экстремум, а именно, минимум в точке
\begin{equation}\notag
  v=(1+\sqrt{3})\frac{\eps_0}{a},
\end{equation}
и тогда из \eqref{eq3_7} получим значение $\eps_1^2=\zeta_3$.
\end{proof}

Две точки пересечения $\dd'_3$ с $\dd_1$ обозначим через $\po_2, \po_3$. Аналогично \eqref{eq3_5} отметим для этих точек зависимость между $\eps_1$ и $\vO$, полученную из \eqref{eq3_7}:
\begin{equation}\label{eq3_8}
\begin{array}{lll}
\po_{2,3}: & \ds \eps_1^2 =\frac{a^3 v^3+5 \eps_0 a^2 v^2+6 \eps_0^2 a v+2\eps_0^3}{a^3 v^2}, &  \ds  v=\vO^2-(\eps_1^2+\frac{\eps_0}{a}).
\end{array}
\end{equation}
Здесь значение $\vO$ взято в точках семейства $\dd'_3$, а в семействе
$\dd_1$ для тех же точек $\po_1,\po_2$ оно будет другим.

Диаграмма Смейла для случая $\eps_1^2 > \zeta_3$ показана на рис.~\ref{fig01},{\it d}.

\section{Показатели Морса и изоэнергетические многообразия}\label{sec5}
Пусть $H_\ell$ есть ограничение функции $H$ на четырехмерное симплектическое многообразие $\mPel$ -- фазовое пространство приведенной системы. Изоэнергетические многообразия $\iso$ -- это уровни функции $H_\ell$, а относительные равновесия на $\mPel$ -- это критические точки $H_\ell$. Поэтому для классификации изоэнергетических многообразий нужно найти индекс Морса функции $H_\ell$ в точках семейств относительных равновесий, описанных в предложении~\ref{prop1}. В свою очередь, индекс Морса функции $H_\ell$ в такой точке равен индексу Морса эффективного потенциала \eqref{eq3_1} как функции на сфере $S^2(\bal)$ в точках \eqref{eq2_6}. Индекс Морса как количество отрицательных собственных чисел второго дифференциала инвариантен относительно выбора систем координат. Однако сами собственные числа, которые ниже для краткости назовем показателями Морса, конечно, неинвариантны. Более того, при использовании избыточных координат для строгого вычисления показателей Морса нужно вводить функции с неопределенными множителями Лагранжа. Этого можно избежать, подобрав подходящим образом дифференциальный оператор, заменяющий второй дифференциал.

\begin{lemma}[\cite{KhRy2011}]\label{lem4}
Пусть функция $f$ определена и дифференцируема в окрестности сферы $S^2(\bal)=\{\bal^2=a^2\}$. Введем дифференциальный оператор
\begin{equation*}
  {\bf D}=\bal{\times}\frac{\partial}{\partial \bal}.
\end{equation*}
Тогда критические точки ограничения $f|_{S^2}$ определяются уравнением
\begin{equation}\notag
  {\bf D} f =0,
\end{equation}
а в качестве показателей для определения индекса Морса можно взять корни квадратного уравнения
\begin{equation}\label{eq4_1}
  \frac{1}{\mu}\det\left({\bf D}^2 f - \mu E\right)=0.
\end{equation}
\end{lemma}

Очевидно, что какие-либо изменения свойств относительных равновесий внутри семейств, объединенных вдобавок по всем значениям физических параметров, могут происходить либо при глобальном изменении семейства на разделяющем значении параметров, либо при пересечении точкой семейства одной из отмеченных выше узловых точек диаграммы Смейла, каковыми являются $T_1,T_2,C,\po_1,\po_2,\po_3$. По договоренности, фиксируем произвольные положительные параметры $a,\eps_0$ и рассматриваем произвольные значения $\eps_1>0$. В расширенном пространстве $\bR^3(\ell,h,\eps_1)$ образы семейств относительных равновесий заполнят поверхности $\dd_i$. Кривые, образованные множествами узловыми точек, разбивают эти поверхности на подобласти, в которых сохраняются свойства соответствующих относительных равновесий. Подобласти будем снабжать вторым индексом, так что они получат обозначения вида $\dd_{ij}$.

Поскольку не все поверхности -- образы относительных равновесий однозначно проектируются на координатные плоскости расширенного пространства, оказывается удобным изображать их на плоскости $(\eps_1,\vO)$, где $\vO$ -- величина, определяющая угловую скорость относительного равновесия, и одновременно параметр кривых в записи \eqref{eq3_2}. Поскольку мы условились считать все параметры и величину $\ell$ неотрицательными, то все зависимости можно записать через $\eps_1^2,\vO^2$. Поэтому в дальнейшем мы применяем обозначения
\begin{equation}\label{eq4_2}
  \zu = \eps_1^2, \qquad \wu=\vO^2,
\end{equation}
используемые в том числе и на иллюстрациях.

\begin{figure}[!ht]
\centering
\includegraphics[width=0.75\textwidth]{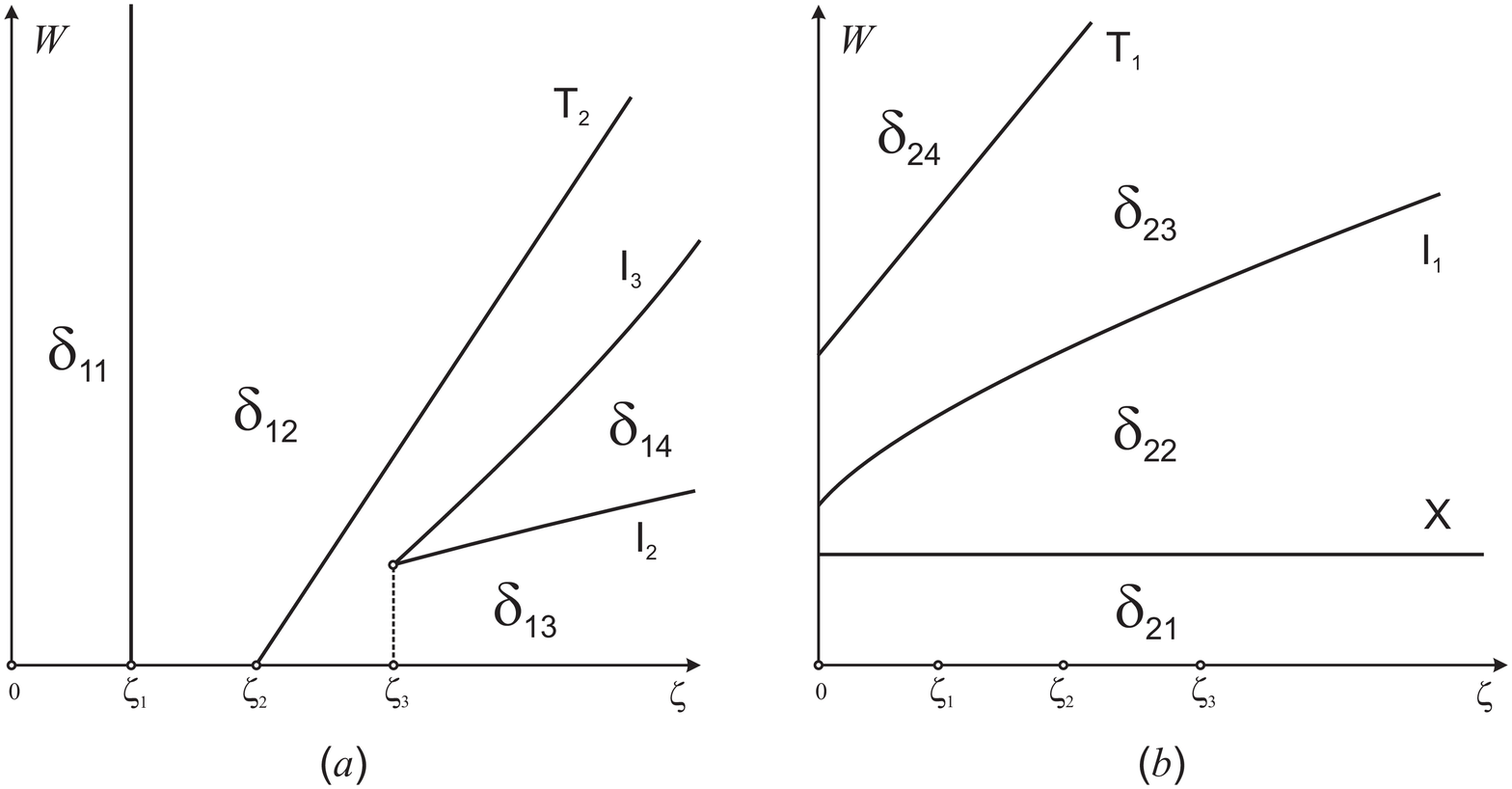}
\caption{Разбиение семейств $(a)$ $\dd_1$; $(b)$ $\dd_2$}\label{figdel12}
\end{figure}

Начнем с семейства $\dd_1$. На плоскости $(\zu,\wu)$ разделяющие кривые отвечают случаям
$\zu=\zeta_1$ (при этом узловыми становятся все точки $\dd_1$, так как из них рождается кривая $\dd_4$), образу точки $T_2$, которая отвечает значению $u=0$ в \eqref{eq3_6} и дает
\begin{equation}\notag
\wu=\zu -\zeta_2,
\end{equation}
и образу пары точек $\po_2,\po_3$, который в силу \eqref{eq3_8} запишется в параметрической форме
\begin{equation}\label{eq4_3}
\zu = \frac{(\eps_0 + a v) (2 \eps_0^2 + 4 a \eps_0 v + a^2 v^2)}{a^3 v^2}, \quad
\wu = \frac{(2 \eps_0^2 + 4 a \eps_0 v + a^2 v^2)^2}{2a v^2 (\eps_0+a v)}, \qquad v>0.
\end{equation}
Здесь $\wu=\vO^2$ отвечает параметру на $\dd_1$ в записи \eqref{eq3_2}. Таким образом, возникают четыре типа относительных равновесий в семействе $\dd_1$. На рис.~\ref{figdel12},{\it a} они обозначены через $\dd_{11}-\dd_{14}$.

Корни уравнения \eqref{eq4_1} в подстановке точек $\dd_1$ из \eqref{eq2_6} имеют вид:
\begin{equation}\notag
\begin{array}{l}
  \mu_1 = a(\eps_0-2\eps_1^2a), \quad \mu_2 = \vO^2+a(\eps_0- \eps_1^2a),
\end{array}
\end{equation}
и знаки показателей Морса на $\dd_1$ таковы: $(+\,+)$ для $\dd_{11}$; $(-\,+)$ для $\dd_{12}$; $(-\,-)$ для $\dd_{13}$ и для $\dd_{14}$. Число минусов есть индекс Морса функции $H_\ell$ в соответствующем относительном равновесии.

Рассмотрим семейство $\dd_2$. На плоскости $(\zu,\wu)$ разделяющие кривые отвечают образу точки $T_1$, которая отвечает значению $v=0$ в \eqref{eq3_4} и дает
\begin{equation}\label{eq4_4}
\wu=a( \eps_0+ a \zu),
\end{equation}
и образу точки $\po_1$, который в силу \eqref{eq3_5} запишется в параметрической форме
\begin{equation}\notag
\zu = \frac{(\eps_0 + a v) (a^2 v^2-2 \eps_0^2)}{a(2\eps_0+a v)^2}, \quad
\wu = \frac{a (2\eps_0^2+2\eps_0 a v + a^2 v^2)^2}{2(\eps_0+av)(2\eps_0+av)^2}, \qquad v \gs \sqrt{2}\frac{\eps_0}{a}.
\end{equation}
Здесь $\wu=\vO^2$ отвечает параметру на $\dd_2$ в записи \eqref{eq3_2}.
Кроме того, на плоскости $(\zu,\wu)$ имеется прямая $\wu=\zeta_1$, соответствующая значению параметра $\vO$ кривой $\dd_2$
\begin{equation}\label{eq4_5}
  \vO = \sqrt{\frac{\eps_0}{2a}},
\end{equation}
которое порождает на $\dd_2$ у всех диаграмм Смейла (см. рис.~\ref{fig01}) точку
\begin{equation}\label{eq4_6}
  X: \quad \ell= a \sqrt{\frac{a \eps_0}{2}}, \quad h=\frac{3 a \eps_0}{2}.
\end{equation}
Об этой точке ниже будет сказано особо при вычислении типов. Подчеркнем, что \eqref{eq4_5} не имеет отношения к разделяющему значению $\zu=\zeta_1$.

Таким образом, возникают четыре типа относительных равновесий в семействе $\dd_2$. На рис.~\ref{figdel12},{\it b} они обозначены через $\dd_{21}-\dd_{24}$.

Корни уравнения \eqref{eq4_1} в подстановке точек $\dd_2$ из \eqref{eq2_6} имеют вид:
\begin{equation}\notag
\begin{array}{l}
  \mu_1 = -a(\eps_0+2\eps_1^2a), \quad \mu_2 = \vO^2-a(\eps_0+ \eps_1^2a).
\end{array}
\end{equation}
Первый всегда отрицателен, второй меняет знак при переходе через точку $T_1$ согласно \eqref{eq4_4}. Итак, знаки показателей Морса на $\dd_2$ таковы: $(-\,-)$ для $\dd_{21} - \dd_{23}$; $(-\,+)$ для $\dd_{24}$, что и определяет индекс Морса приведенного гамильтониана $H_\ell$.

\begin{figure}[!ht]
\centering
\includegraphics[width=0.35\textwidth]{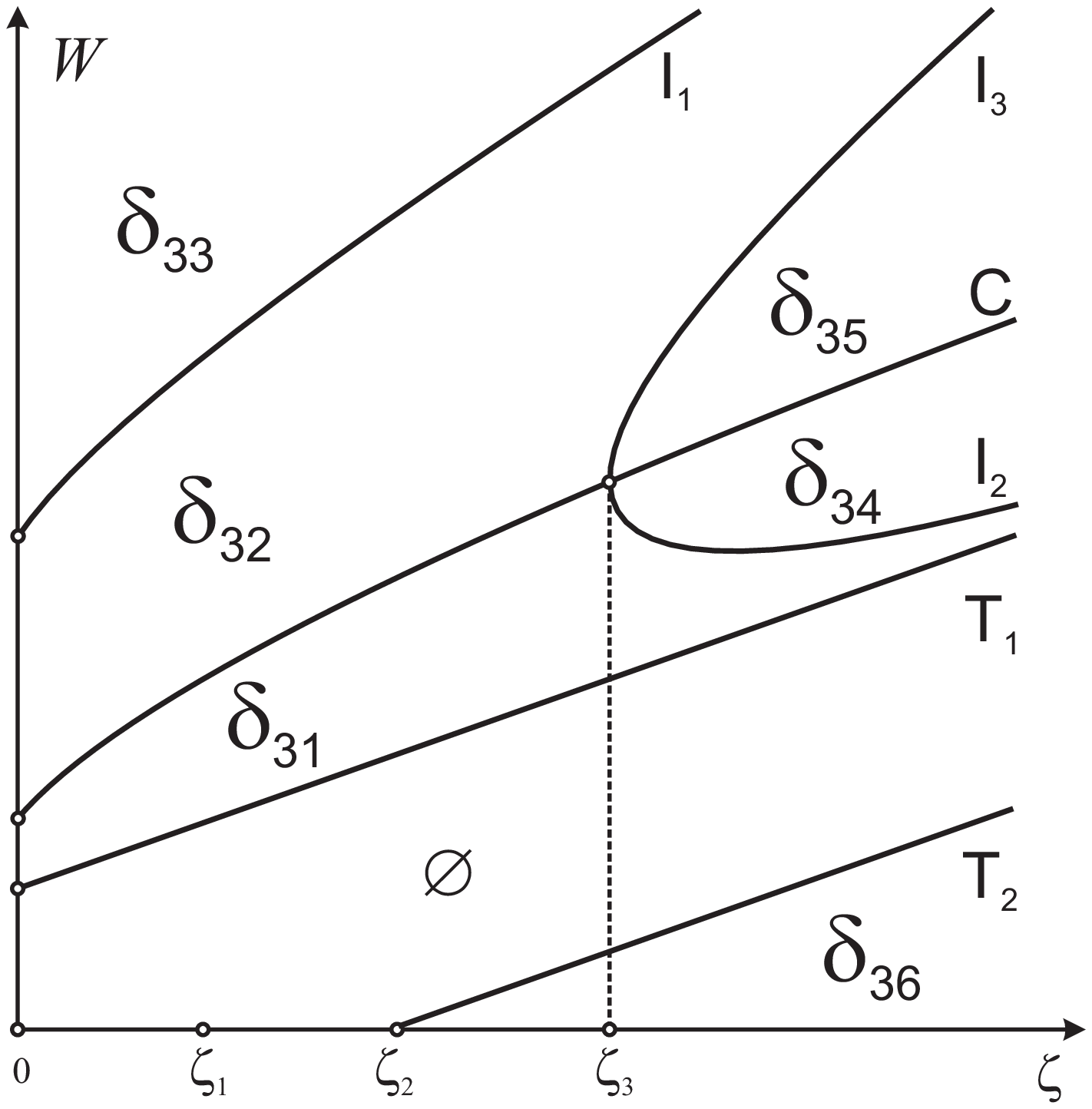}
\caption{Разбиение семейства $\dd_3$}\label{figdel3}
\end{figure}

Обратимся к семейству $\dd_3$. На плоскости $(\zu,\wu)$ разделяющие кривые -- это образ точек $T_1, T_2$, которые отвечают значению $v=0$ в \eqref{eq3_4} и значению $u=0$ в \eqref{eq3_6}. Используя параметр $\wu=\vO^2$ семейства $\dd_3$, получим
\begin{equation}\notag
\begin{array}{l}
\ds  T_1: \wu = \zu + \zeta_2, \qquad T_2: \wu = \zu - \zeta_2.
\end{array}
\end{equation}
Из \eqref{eq3_5} найдем выражения для образов точки возврата $C$ и точки пересечения $\po_1$:
\begin{equation}\notag
\begin{array}{l}
C: \quad \ds \zu= \frac{a^3 v^3  + 3 \eps_0 a^2 v^2 -2 \eps_0^3}{4 a \eps_0^2},\;
\ds  \wu =\frac{(\eps_0 + a v) (2 \eps_0^2 + 2 \eps_0 a v + a^2 v^2)}{4 a \eps_0^2}, \quad  v\gs (\sqrt{3}-1)\frac{\eps_0}{a};\\
\po_1: \quad\ds \zu = \frac{(a v+\eps_0 ) (a^2 v^2-2 \eps_0^2 )}{a (a v+2 \eps_0)^2}, \quad
\wu =\frac{2 (\eps_0 + a v)^3}{a (2 \eps_0 + a v)^2}, \quad v \gs \sqrt{2}\frac{\eps_0}{a}.
\end{array}
\end{equation}
Наконец, образ пары точек $\po_2,\po_3$ найдем из \eqref{eq3_8}:
\begin{equation}\notag
\po_2, \po_3: \quad \zu = \frac{(\eps_0 + a v) (2 \eps_0^2 + 4 a \eps_0 v + a^2 v^2)}{a^3 v^2}, \;
\wu = \frac{2 (\eps_0 + a v)^3}{a^3 v^2}, \quad v>0.
\end{equation}
Отметим, что, в отличие от представления \eqref{eq4_3}, здесь $\wu=\vO^2$ -- параметр семейства $\dd_3$.

Нанесем полученные кривые на плоскость $(\zu,\wu)$, получим разбиение семейства $\dd_3$ на шесть подсемейств, как показано на рис.~\ref{figdel3}. Между прямыми, порожденными точками $T_1,T_2$, относительных равновесий нет, что соответствует разрыву между подсемействами $\dd'_3$ и $\dd''_3$.

Корни уравнения \eqref{eq4_1} в подстановке точек $\dd_3$ из \eqref{eq2_6} простого выражения не имеют. Для того чтобы установить индекс Морса функции $H_\ell$ на семействе $\dd_3$, выпишем вначале свободный член уравнения \eqref{eq4_1}. Отбрасывая заведомо положительные сомножители, получим
\begin{equation}\notag
\begin{array}{l}
\sgn \mu_1 \mu_2 =\sgn  \left\{[a^2(\zu - \wu)^2- \eps_0^2]
[a^2 (\wu - \zu)^3   - \eps_0^2(\zu + 3 \wu)]\right\}.
\end{array}
\end{equation}
Первый сомножитель в правой части обращается в нуль в точках $T_1,T_2$, являющихся граничными для семейств $\dd'_3,\dd''_3$, то есть внутри семейств влияния на изменение знаков $\mu_1,\mu_2$ он не оказывает. Более того, он положителен внутри обоих семейств. Обращение в нуль второго сомножителя соответствует, как легко видеть, точке возврата $C$. Поэтому ниже кривой, порожденной точкой $C$ этот сомножитель отрицателен (например, при $\wu=0$), а выше -- положителен (что очевидно при больших $\wu > 0$). Итак, $\mu_1 \mu_2<0$ на $\dd_{31}, \dd_{34},\dd_{36}$ и $\mu_1 \mu_2>0$ на $\dd_{32}, \dd_{33},\dd_{35}$. Осталось определить общий знак $\mu_1,\mu_2$ для последних трех подсемейств.
Запишем выражение для суммы показателей:
\begin{equation*}
\begin{array}{l}
\ds  \mu_1+\mu_2=a^2(\zu-2\wu)+\frac{1}{(\wu - \zu)^2 [\eps_0^2 \wu + a^2 (\wu - \zu)^2 (\wu + 2 \zu)]}{\times}\\[3mm]
\qquad {\times} \left[ a^2 \eps_0^2 (\wu - \zu)^2 (5 \wu^2 + 3 \wu \zu - 3 \zu^2)-a^4 (\wu - \zu)^4 \zu^2-\eps_0^4 \wu (3 \wu + 5 \zu)\right] = \\[3mm]
  \qquad = a^2(\zu-2\wu) + O(\wu^{-1}).
\end{array}
\end{equation*}
Очевидно, при больших $\wu>0$ эта сумма отрицательна, а так как выше кривой, отвечающей точке возврата $C$, знаки $\mu_1,\mu_2$ неизменны, то на $\dd_{32}, \dd_{33},\dd_{35}$ оба показателя отрицательны.

Наконец, образ на плоскости $(\ell,h)$ семейства $\dd_4$, существующей при условии $\eps_1^2 > \zeta_1$, не имеет никаких пересечений с другими кривыми, поэтому ни семейство, ни обозначенная так же кривая, дополнительных разбиений не получают. Корни уравнения \eqref{eq4_1} в подстановке точек $\dd_4$ из \eqref{eq2_6} таковы:
\begin{equation}\notag
\begin{array}{l}
\ds \mu_1 = \frac{4\eps_1^2 a^2-\eps_0^2}{2\eps_1^2}, \quad
\mu_2 = \frac{8\eps_1^2a^4(\eps_1^2 + \vO^2)^2}{\eps_0^2+4\eps_1^2a^2(\eps_1^2+2\vO^2)}.
\end{array}
\end{equation}
Очевидно, в области существования они всегда положительны, так что здесь положительны все показатели Морса функции $H_\ell$.


Найденная информация об относительных равновесиях собрана в табл.~1 (последний столбец обсуждается в следующем параграфе). Зная количество точек и индекс Морса, находим топологические типы $\iso$ и характер бифуркаций (в таблице они записаны в направлении возрастания $h$). В расширенном пространстве $\bR^3(\ell,h,\eps_1)$ объединение диаграмм Смейла определяет семь областей с непустыми $\iso$ (очевидно, что ниже всех парабол $\iso=\varnothing$). Эти области обозначены через $\mtA - \mtG$. В случае рис.~\ref{fig01},{\it a} все эти многообразия известны из случая Ковалевской \cite{Jacob}: $\mtA$)~$S^3$; $\mtB$)~$\bR P^3$; $\mtC$)~$S^2{\times}S^1$; $\mtD$)~$N^3_2$. Здесь и далее через $N^3_m$ обозначена связная сумма $m$ экземпляров $S^2{\times}S^1$. Такое изоэнергетическое многообразие получается в схеме Смейла как приведенное расслоение окружностей над двумерным диском с $m$ дырками. В частности, естественно считать $N^3_0=S^3$, $N^3_1 =S^2{\times}S^1$. Для области $\mtE$ получим $\iso=2S^3$ из очевидной бифуркации на $\dd_4$ (рис.~\ref{fig01},{\it b}). В область $\mtF$ удобно спуститься из области $\mtA$ через точку семейства $\dd_{13}$ с индексом Морса 2, что соответствует вырезанию дырки в диске на сфере $S^2(\bal)$, отвечающем области $\mtA$. Поэтому многообразие $\iso$ в области $\mtF$ связно и диффеоморфно $S^2{\times}S^1$. В частности, переход к нему из области $\mtE$ через пару точек семейства $\dd_{36}$ -- это приклейка двух ручек к двум сферам так, что результат связен, то есть хотя бы одна ручка приклеивается своими концами к разным сферам. Наконец, в область $\mtG$ также удобно спуститься из области $\mtD$ через точку семейства $\dd_{14}$ с индексом Морса 2, что соответствует вырезанию дырки в диске на сфере $S^2(\bal)$, отвечающем области $\mtD$, который уже имеет две дырки. Поэтому результатом для $\mtG$ является $N^3_3$. Интересно отметить, что изоэнергетическое многообразие $N^3_3$ в классической динамике твердого тела (движение вокруг неподвижной точки в поле только силы тяжести) возможно в случае общего положения центра масс \cite{Gash2003} (историю вопроса можно найти в \cite{BorMam1}), однако, в интегрируемых задачах оно ранее появлялось лишь в случаях Клебша \cite{Cleb} и Соколова \cite{So2001} для задачи Кирхгофа движения тела в жидкости (что в соответствующей задаче о движении вокруг неподвижной точки в первом случае означает наличие центрального ньютоновского поля вместо поля силы тяжести, а во втором предполагает наличие гироскопических сил, зависящих от ориентации тела). Изоэнергетические многообразия случая Клебша классифицированы в работе \cite{Oshem} (где, собственно, впервые и обнаружилось изоэнергетическое многообразие $N^3_3$), фазовая топология случая Соколова изучена в работе \cite{Ry2003}. Случай Соколова для уравнений Кирхгофа является предельным для рассматриваемой здесь задачи при $\eps_0 \to 0$. При таком переходе сохраняются области $\mtB,\mtC,\mtE,\mtF,\mtG$, а новых областей не возникает. В частности, любая из областей $\mtA - \mtG$ при деформации параметров имеет выход либо на классический случай Ковалевской ($\eps_1=0$), либо на случай Соколова для уравнений Кирхгофа ($\eps_0=0$).

\def\rul{\rule[-9pt]{0pt}{22pt}}

{

\centering
\small
\begin{longtable}{|c|c|c|c|c|c|}
\multicolumn{6}{r}{\fts{Таблица 1. Классы относительных равновесий}}\\
\hline
\fns{Код} & \fns{\begin{tabular}{c}К-во\\[-3pt] точек\end{tabular}} & \fns{\begin{tabular}{c}Индекс\\[-3pt] Морса\end{tabular}} & Сегменты &Бифуркация $\iso$& Тип\\
\hline\endfirsthead%
\multicolumn{6}{r}{\fts{Таблица 1. Продолжение}}\\
\hline
\fns{Код} & \fns{\begin{tabular}{c}К-во\\[-3pt] точек\end{tabular}} & \fns{\begin{tabular}{c}Индекс\\[-3pt] Морса\end{tabular}} & Сегменты &Бифуркация $\iso$& Тип\\
\hline\endhead
\rul $\dd_{11}$ & 1 & 0 & $[A_1,\infty)$ & $\varnothing \to S^3$& ``центр-центр'' \\
\hline
\rul $\dd_{12}$ & 1 & 1 & \begin{tabular}{l}$[A_1,\infty), \; (b)$\\$[T_2,\infty), \; (c,d)$\end{tabular} &{$2S^3\to S^3\#S^3=S^3$}& ``центр-седло'' \\
\hline
\rul $\dd_{13}$ & 1 & 2 & \begin{tabular}{l}$[A_1,T_2], \; (c)$\\$[A_1,\po_3]\cup [\po_2,\infty), \; (d)$\end{tabular} &{$S^2{\times}S^2 \to S^3$}& ``седло-седло'' \\
\hline
\rul $\dd_{14}$ & 1 & 2 & $[\po_3,\po_2],\; (d)$ & $N^3_3\to N^3_2$& ``седло-седло'' \\
\hline
\rul $\dd_{21}$ & 1 & 2 & $[A_2,X]$ & $S^3\to \bR P^3$& ``седло-седло'' \\
\hline
\rul $\dd_{22}$ & 1 & 2 & $[X,\po_1]$ & $S^3\to \bR P^3$& ``седло-седло'' \\
\hline
\rul $\dd_{23}$ & 1 & 2 & $[\po_1,T_1]$ & $N^3_2\to S^2{\times}S^1$& ``седло-седло'' \\
\hline
\rul $\dd_{24}$ & 1 & 1 & $[T_1,\infty)$ & $S^3\to S^2{\times}S^1$& ``центр-седло'' \\
\hline
\rul $\dd_{31}$ & 2 & 1 & \begin{tabular}{l}$[T_1,C], \; (a,b,c)$\\$[T_1,\po_2], \; (d)$\end{tabular} &$S^3\to N^3_2$& ``центр-седло'' \\
\hline
\rul $\dd_{32}$ & 2 & 2 & \begin{tabular}{l}$[C,\po_1], \; (a,b,c)$\\$[\po_3,\po_1], \; (d)$\end{tabular} &$N^3_2\to S^3$& ``центр-центр'' \\
\hline
\rul $\dd_{33}$ & 2 & 2 & $[\po_1,\infty)$ & $S^2{\times}S^1\to \bR P^3$& ``центр-центр'' \\
\hline
\rul $\dd_{34}$ & 2 & 1 & $[\po_2,C],\;(d)$ & $S^2{\times}S^1\to N^3_3 $& ``центр-седло'' \\
\hline
\rul $\dd_{35}$ & 2 & 2 & $[C,\po_3],\;(d)$ & $N^3_3 \to S^2{\times}S^1$& ``центр-центр'' \\
\hline
\rul $\dd_{36}$ & 2 & 1 & $[A_3,T_2],\;(c,d)$ & $2S^3\to S^2{\times}S^1$& ``центр-седло'' \\
\hline
\rul $\dd_{4}$ & 2 & 0 & $[A_4,\infty),\;(b,c,d)$ & $\varnothing \to 2S^3$& ``центр-центр'' \\
\hline
\end{longtable}

}


\section{Типы и устойчивость относительных равновесий}\label{sec6}
Хорошо известно \cite{Malk}, что в неподвижной точке матрица линеаризации канонических уравнений с гамильтонианом $H$ задает оператор $A_H: \bR^6 \to \bR^6$, у которого характеристический многочлен содержит только четные степени. Соответствующие собственные числа полностью определяют характер устойчивости, если все они различны. В силу вырожденности скобки \eqref{eq1_1} или, что то же самое, в силу наличия интегралов \eqref{eq1_4}, два собственных числа оператора $A_H$ нулевые. Обозначим через $\roo_H(\mu)$ характеристический многочлен $A_H$, сокращенный на $\mu^2$. Очевидно, многочлен $\roo_H(\mu)$ есть биквадрат. Заменяя в уравнениях \eqref{eq1_2} гамильтониан $H$ на первый интеграл $K$, той же процедурой получим оператор $A_K$ и биквадратный трехчлен $\roo_K(\mu)$.

Полагая $\ld=\mu^2$ и используя формулы \eqref{eq2_6} выпишем явно квадраты корней $\roo_H(\mu)$ в точках найденных семейств относительных равновесий:
\begin{equation}\notag
\begin{array}{l}
\ds  \dd_1: \quad \ld_1=a^2\left(\eps_1^2-\frac{\eps_0}{2a}\right), \quad \ld_2= a^2\left[(\eps_1^2-\frac{\eps_0}{a})-\vO^2\right];\\[3mm]
\ds  \dd_2: \quad \ld_1 = a^2\left(\eps_1^2+\frac{\eps_0}{2a}\right), \quad
\ld_2= a^2\left[(\eps_1^2+\frac{\eps_0}{a})-\vO^2\right];\\[3mm]
\ds \dd_3: \quad \left\{
\begin{array}{l}
\ds  \ld_1 = -\frac{a^2(\eps_1^2+\vO^2)}{(\eps_1^2-\vO^2)^2}\left[(\eps_1^2-\vO^2)^2-\frac{\eps_0^2}{a^2}\right] \\[3mm]
\ds  \ld_2 = \frac{a^2}{4(\eps_1^2-\vO^2)^2}\left[(\eps_1^2-\vO^2)^3+\frac{\eps_0^2}{a^2}(\eps_1^2+3\vO^2)\right]
\end{array} \right. ;\\[3mm]
\ds \dd_4: \quad \ld_1= -\frac{a^2}{\eps_1^2} \left[\eps_1^4-\frac{\eps_0^2}{4a^2}\right], \quad \ld_2= -a^2(\eps_1^2+\vO^2).
\end{array}
\end{equation}

Напомним, что в гамильтоновой системе с двумя степенями свободы, имеющей два функционально независимых первых интеграла $H,K$, тип неподвижной точки определяется для так называемых  невырожденных точек. Критерий невырожденности состоит в том, что операторы $A_H,A_K$ линейно независимы и найдется такая их линейная комбинация, у которой все собственные числа различны \cite{LerUmI,BolFom}. Пусть $A$ такая комбинация. Оператор $A$ называется в этом случае регулярным элементом (алгебры симплектических операторов, порожденной парой $A_H,A_K$). Говорят, что неподвижная точка имеет тип ``центр-центр'', если все собственные числа $A$ чисто мнимые, тип ``седло-седло'', если все они вещественные, и тип ``центр-седло'', если одна пара собственных чисел чисто мнимая, а вторая вещественна. Теоретически имеется еще один случай, когда собственные числа $A$ имеют вид $(\pm a \pm b \ri)$ с $ab \ne 0$. Такие неподвижные точки называются фокусными. При этом характеристический многочлен регулярного элемента неприводим над $\bR$. Однако уже из полученных выражений для $\roo_H(\mu)$ следует, что в рассматриваемой задаче фокусных точек нет.

Найденные значения показывают, что многочлен $\roo_H(\mu)$ имеет нулевые корни только в некоторых из случаев, отмеченных выше как разделяющие для семейств $\dd_{i}$. Более точно, нулевые корни существуют
\begin{itemize}
\item в семействе $\dd_1$ лишь при значении $\eps_1^2=\zeta_1$, когда от $\dd_1$ отщепляется семейство $\dd_4$, и в точке $T_2$;
\item в семействе $\dd_2$ только в точке $T_1$;
\item в семействе $\dd_3$ только в точках $T_1, T_2$ и $C$.
\end{itemize}

В семействе $\dd_4$, за исключением момента его появления, все корни $\roo_H(\mu)$ чисто мнимые.

Рассмотрим возможность другого типа вырождения, а именно, случай наличия двух одинаковых пар корней $\ld_1=\ld_2$. На кривых $\dd_1,\dd_4$ это, очевидно, невозможно. На кривой $\dd_2$ это равенство дает значение \eqref{eq4_5}, которому отвечает точка \eqref{eq4_6} на диаграммах Смейла. При этом совпадающие пары  вещественны. Вычисляя многочлен $\roo_\Phi(\mu)$ для произвольной комбинации $\Phi=\vk_1 K + \vk_2 H$, убеждаемся, что совпадение пар корней сохраняется для относительного равновесия в прообразе точки $X$ для любого такого интеграла $\Phi$. Поэтому это относительное равновесие всегда вырождено, хотя на характер бифуркации $\iso$ это не влияет, и по-прежнему касательное пространство к $\mPel$ можно разложить в прямую сумму плоскостей, в проекции на каждую из которых соответствующая неподвижная точка приведенной системы будет выглядеть седловой. Заметим, что операторы $A_H$ и $A_K$ остаются в этой точке линейно независимыми.

На кривой $\dd_3$ в обозначениях \eqref{eq4_2} равенство $\ld_1=\ld_2$ дает
\begin{equation}\notag
  a^2(\wu-\zu)^2(3\wu+5\zu)-\eps_0^2(\wu+3\zu)=0.
\end{equation}
Выполним подстановку $\wu=\zu+\zeta_2 x$. Из условий для параметров на кривой $\dd_3$
следует, что $|x| \gs 1$. Получим
\begin{equation*}
  \zu = \frac{\eps_0}{4a} \frac {(3x^2-1)x}{2x^2-1}, \quad \wu = -\frac{\eps_0}{4a} \frac {(5x^2-3)x}{2x^2-1},
\end{equation*}
то есть $\zu,\wu$ должны иметь разные знаки, что невозможно. Итак, на кривой $\dd_3$ случаев $\ld_1=\ld_2$ нет.

В результате приходим к следующему утверждению.

\begin{propos}\label{prop4}
В интегрируемом случае Ковалевской\,--\,Соколова все относительные равновесия невырождены, за исключением тех, которые лежат в прообразе точек $T_1$, $T_2$, $C$, $X$, и всех относительных равновесий совпадающих семейств $\dd_1=\dd_4$ при разделяющем значении параметров $\eps_1^2=\zeta_1$. Невырожденные относительные равновесия имеют тип, указанный в последнем столбце табл.~$1$.
\end{propos}

Поскольку все типы относительных равновесий удалось определить, исходя из многочлена $\roo_H(\mu)$, то они полностью определяют характер устойчивости соответствующих неподвижных точек приведенной системы: точки типа ``центр-центр''  устойчивы по всем переменным, точки типа ``седло-седло''  неустойчивы по всем переменным, а точки типа ``центр-седло''  устойчивы по двум переменным, а по двум -- неустойчивы.

Топология так называемой насыщенной четырехмерной окрестности неподвижной точки интегрируемой системы с двумя степенями свободы полностью определена типом для точек ``центр-центр'' и ``центр-седло'' \cite{LerUmI,BolFom}. Для седловых особенностей возможны различные варианты. Их определение требует знания топологии двумерных регулярных уровней первых интегралов приведенных систем в окрестности неподвижных точек и характера их бифуркаций. Следующий раздел содержит необходимую информацию по
грубой топологии рассматриваемой задачи.


\section{Разделение переменных и дискриминантные поверхности}\label{sec7}
Система на коалгебре $\mg_0$ со скобкой \eqref{eq1_1} и гамильтонианом \eqref{eq1_3} тесно связана с обобщением случая Ковалевской на пучок коалгебр $\mg_\vk = \{(\bm,\bp)\}$ ($\vk \ne 0$), в котором скобка Пуассона имеет вид
\begin{equation}\label{eq6_1}
  \{M_i,M_j\}=\eps_{ijk}M_k, \quad \{M_i,\vvp_j\}=\eps_{ijk}\vvp_k, \quad \{\vvp_i,\vvp_j\}=\vk \eps_{ijk} M_k.
\end{equation}
Это обобщение найдено в работе \cite{KomarTMF1981}, там же предложена цепочка замен переменных, приводящих в конечном итоге к разделенным уравнениям типа Ковалевской. Более подробно этот результат представлен в работе \cite{KomKuz1990}. Во многих публикациях имеются модификации разделения переменных для этой задачи (см., например, \cite{Jur, BorMamKholm,Drag,Drag2}). Формулы построения разделения переменных в \cite{Drag,Drag2}, по существу, следуют тому варианту, который предложил Кёттер \cite{Kott} для классического случая Ковалевской, однако, полученные переменные разделения все же не коммутируют. Чтобы записать результат разделения для рассматриваемой здесь задачи, рассмотрим диффеоморфизм коалгебры $\mg_0$ на коалгебру $\mg_\vk$ с $\vk<0$, которая, как известно, есть $so(3,1)^*$, причем нормировкой вектора $\bp$ всегда можно добиться равенства $\vk=-1$. Этот пуассонов диффеоморфизм $(\bm,\bal) \mapsto (\bm,\bp)$, переводящий скобку \eqref{eq1_1} в скобку \eqref{eq6_1} в общем случае имеет вид \cite{KoTsSo2003}
\begin{equation}\label{eq6_2}
  \bp = c_1 \bal + c_2 \bal{\times}\bm, \qquad c_1,c_2 ={\rm const}.
\end{equation}
Очевидно, для того чтобы гамильтониан \eqref{eq1_3} перешел в гамильтониан Ковалевской
\begin{equation}\notag
  H=\frac{1}{4}(M_1^2+M_2^2+2M_3^2)- c \vvp_1,
\end{equation}
необходимо положить
\begin{equation}\label{eq6_3}
  c_1=c^{-1}\eps_0,\quad c_2=c^{-1}\eps_1.
\end{equation}
Преобразование, обратное к \eqref{eq6_2}, запишется в виде
\begin{equation}\label{eq6_4}
\begin{array}{rcl}
\bal &=& \ds\frac{c}{\eps_0(\eps_0^2+\eps_1^2 \bm^2)}\Bigl[ \eps_0^2 \bp + \eps_0 \,\eps_1 \bm{\times}\bp +\eps_1^2 (\bp \cdot \bm) \bm \Bigr] =   \\[3mm]
{}  &=& \ds \frac{1}{\eps_0}\left(c\, \bp+\ds\frac{\eps_1 c}{\eps_0^2+\eps_1^2 \bm^2} \Bigl[\eps_0 \bm{\times}\bp+\eps_1\bm{\times}( \bm{\times}\bp) \Bigr]\right).
\end{array}
\end{equation}
Из \eqref{eq6_2}, \eqref{eq6_3} для параметра пучка найдем
\begin{equation*}
  \vk =-\frac{a^2\eps_1^2}{c^2},
\end{equation*}
поэтому нужная нормировка скобки на $so(3,1)^*$ достигается выбором параметра $c$. Функции Казимира для коалгебры $\mg_{-1}$ возьмем в виде
\begin{equation}\notag
  G=\bp^2-\bm^2, \qquad F=\frac{1}{2} \bm \cdot \bp.
\end{equation}
Соответствующие константы обозначим через $g,f$. Преобразование \eqref{eq6_4} с учетом нормировки связывает константы $c,f,g$ и $\eps_1, a, \ell$ соответствием
\begin{equation}\notag
  c =\eps_1 a, \quad g=\frac{\eps_0^2a^2-4\eps_1^2\ell^2}{\eps_1^2a^2}, \quad f=\frac{\eps_0 \ell}{\eps_1 a}.
\end{equation}
Обратное к нему
\begin{equation}\notag
  \eps_1=\frac{\sqrt{2}\eps_0}{\sqrt{\sqrt{g^2+16f^2}+g}} , \quad a=\frac{c}{\sqrt{2}\eps_0} \sqrt{\sqrt{g^2+16f^2}+g}, \quad \ell =\frac{c f}{\eps_0}.
\end{equation}
Здесь, как и ранее, удобно считать $\eps_0 > 0$ фиксированным.

Таким образом, все известные разделения переменных для случая Ковалевской\,--\,Ко\-ма\-ро\-ва на $so(3,1)^*$ переносятся на задачу Ковалевской\,--\,Соколова. Полученные в результате зависимости допускают предельный переход при $\eps_1 \to 0$ к классическому волчку Ковалевской (как и случай Ковалевской\,--\,Ко\-ма\-ро\-ва, если не нормировать параметр $\vk$). С другой стороны, сам волчок Ковалевской\,--\,Соколова является предельным для обобщенного двухполевого гиростата Соколова\,--\,Цыганова, интегрируемость которого доказана в \cite{SoTs2002} (при обращении в ноль постоянного гиростатического момента и второго однородного поля). Для полного согласования с исследованиями особых (критических) движений классического волчка Ковалевской \cite{Appel,Ipat,KhPMM83} и гиростата Соколова\,--\,Цыганова \cite{Ry2013,KhRCD2014} будем строго следовать методу Кёттера.

Введем переменные ($\ri^2=-1$)
\begin{equation}\label{eq6_5}
\begin{array}{lll}
\ds  w_1 = \frac{1}{2} (M_1 +\ri M_2), &  \ds w_2 = \frac{1}{2} (M_1 - \ri M_2), & w_3 = M_3, \\[2mm]
  x_1 =\va_1+\ri \va_2, & x_2 =\va_1-\ri \va_2, & z=\va_3,
\end{array}
\end{equation}
и обозначим
\begin{equation}\notag
\begin{array}{l}
R(w)=-w^4+2hw^2+4\eps_0 \ell w-k+\eps_1^2[a^2(2h+\eps_1^2a^2)-4\ell^2]+\eps_0^2a^2,\\[2mm]
R(w_1,w_2)=\dfrac{1}{2}[R(w_1)+R(w_2)+(w_1^2-w_2^2)^2].
\end{array}
\end{equation}
Прямое применение метода Кёттера приводит к следующему результату.

\begin{propos}[\cite{KoTsSo2003}]\label{prop5}
Переменные типа Ковалевской
\begin{equation}\notag
\begin{array}{l}
\ds s_1=\frac{R (w_1,w_2)-\sqrt{R(w_1)R(w_2)}}{(w_1-w_2)^2},\quad s_2=\frac{R(w_1,w_2)+\sqrt{R(w_1)R(w_2)}}{(w_1-w_2)^2}
\end{array}
\end{equation}
коммутируют $\{s_1,s_2\}=0$ и их динамика описывается разделенными уравнениями
\begin{equation}\notag
\begin{array}{l}
\ds{(s_1-s_2)^2\dot s_1^2=-2P(s_1)\varphi(s_1),\quad (s_1-s_2)^2\dot s_2^2=-2P(s_2)\varphi(s_2),}
\end{array}
\end{equation}
где
\begin{equation}\notag
\begin{array}{l}
P(s)=[s-(h+\eps_1^2a^2)]^2-k, \\[2mm]
\varphi(s)=s^3-2hs^2+[(h+\eps_1^2a^2)^2+a^2\eps_0^2- 4\eps_1^2\ell^2-k]s-2\eps_0^2\ell^2.
\end{array}
\end{equation}
\end{propos}

Оказывается, что в этой задаче бифуркации первых интегралов в точности соответствуют дискриминантному множеству многочлена, участвующего в разделенных уравнениях, с обычными оговорками вещественности решений.

\begin{theorem}\label{theo1}
Бифуркационная диаграмма полного отображения момента
\begin{equation}\notag
\mathcal{F}=H{\times}L{\times}K: \mP^5 \to \bR^3
\end{equation}
случая Ковалевской\,--\,Соколова есть отвечающая вещественным решениям часть дискриминантного множества многочлена $S(s)=P(s) \varphi(s)$, которое состоит из следующих поверхностей:
\begin{eqnarray}
& & \ds \Pi_1: \quad k=0; \nonumber\\
& & \ds \Pi_2: \quad k =\frac{1}{4\eps_1^4}\left[2\eps_1^2(h+\eps_1^2a^2)+\eps_0^2 \right]^2;  \nonumber\\
& & \ds \Pi_3: \quad k=\frac{[a^2(h+\eps_1^2a^2)-2\ell^2]^2}{a^4}; \nonumber\\
& & \ds \Pi_4: \quad \left\{\begin{array}{l}
\ds{\ell^2=\frac{1}{\eps_0^2}(h-s)s^2}\\[5mm]
\ds{k=\frac{4\eps_1^2}{\eps_0^2}s^3+\left(3-\frac{4\eps_1^2}{\eps_0^2}h\right)s^2-4hs+
a^2\eps_0^2+(h+\eps_1^2a^2)^2}
\end{array}\right. \nonumber.
\end{eqnarray}
\end{theorem}

Доказательство можно получить из результатов работ \cite{Ry2013,KhRCD2014}, где дано описание бифуркационных диаграмм и множества критических точек отображения момента для гиростата в двойном поле. В частности, в этих работах множество критических точек отображения момента представлено как объединение так называемых критических подсистем.

\section{Критическое множество и типы критических точек}\label{sec8}
Чтобы сформулировать описание критического множества отображения момента в случае Ковалевской\,--\,Соколова, напомним понятие критической подсистемы \cite{KhRCD05, KhND07, KhRCD2014}. Рассмотрим неприводимую интегрируемую систему с тремя степенями свободы с инволютивным набором интегралов, который обозначим так же, как и в нашей задаче, через $H,L,K$. Предположим, для простоты, что в ней отсутствуют фокусные особенности ранга~1. Тогда в представлении бифуркационной диаграммы отображения момента в виде двумерного клеточного комплекса, будут отсутствовать изолированные одномерные клетки. Пусть
\begin{equation}\notag
\Phi (h,\ell,k)=0
\end{equation}
уравнение двумерной поверхности, несущей на себе некоторую двумерную клетку бифуркационной диаграммы. Рассмотрим систему уравнений
\begin{equation}\label{eq7_1}
\Phi=0, \quad d\Phi=0,
\end{equation}
в которой вместо $(h,\ell,k)$ подставлены соответствующие функции. Ранг этой системы на множестве ее решений $\mcm_\Phi$ почти всюду равен двум. Поэтому множество решений есть замыкание четырехмерного многообразия $\mcm^0_\Phi$, состоящего целиком из критических точек ранга~2. На $\mcm_\Phi$ индуцируется динамическая система, которая почти всюду гамильтонова с двумя степенями свободы. Множество $\mcm_\Phi$ называем критической подсистемой, порожденной функцией $\Phi$.

С другой стороны многообразие $\mcm^0_\Phi$ может быть записано в виде системы инвариантных соотношений
\begin{equation}\label{eq7_2}
  \psi_1 =0, \qquad \psi_2 =0,
\end{equation}
в которой функции $\psi_1,\psi_2$ независимы на $\mcm^0_\Phi$. В таком представлении можно говорить, что критическая подсистема $\mcm_\Phi=\mathrm{Cl}(\mcm^0_\Phi)$ определена инвариантными соотношениями \eqref{eq7_2}.

Известно \cite{FomSimGeom}, что 2-форма, индуцированная на $\mcm^0_\Phi$ симплектической структурой фазового пространства, вырождается на множестве $\{\psi_1,\psi_2\}=0$.

\begin{remark}\label{rem1}
{\small
Обычно систему инвариантных соотношений можно выбрать так, чтобы было
\begin{equation}\notag
  \dot \psi_1 = \vk_1 \psi_2, \qquad \dot \psi_2 = \vk_2 \psi_1
\end{equation}
с некоторыми функциями $\vk_1,\vk_2$ (точкой обозначено дифференцирование в силу исходной гамильтоновой системы). Тогда (см., например, {\rm \cite{BogRus2}}) скобка $\{\psi_1,\psi_2\}$ будет частным интегралом на $\mcm^0_\Phi$ и, следовательно, на его замыкании $\mcm_\Phi$.
В {\rm \cite{BolFom}} отмечено, что на многообразиях, сформированных невырожденными критическими точками одного ранга, индуцированная симплектическая структура не вырождается. В частности, это означает, что нулевой уровень частного интеграла $\{\psi_1,\psi_2\}$ целиком состоит из вырожденных критических точек.
}
\end{remark}

В нашем случае механическая система на $SO(3)$ приводима и, после факторизации по группе симметрий $S^1$ вращений трехмерного пространства вокруг вектора $\bal$, решения систем вида \eqref{eq7_1} заполняют трехмерные многообразия в $\mP^5$, расслоенные уровнями интеграла $L$ на почти гамильтоновы системы с одной степенью свободы в $\mPel$. В частности, рангом критической точки отображения момента будем называть ее ранг в приведенной системе на $\mPel$.

Далее следуем процедуре, описанной в \cite{KhRCD2014}, с учетом приводимости нашей системы. Пусть $\Phi$ -- функция, порождающая критическую подсистему $\mcm_\Phi$. Рассмотрим композицию $\Phi \circ\mathcal{F}$ функции $\Phi$ с отображением момента на всем $\bR^6(\bm,\bal)$ и вычислим характеристический многочлен соответствующего этой функции симплектического оператора, то есть оператора, задающего линеаризацию гамильтонова векторного поля $\sgrad (\Phi\circ\mathcal{F})$. Допуская очевидную вольность, обозначим оператор через $A_\Phi$ (строго говоря, здесь в индексе следовало бы указывать $\Phi\circ\mathcal{F}$). Он имеет четыре нулевых собственных значения. Поэтому здесь через $\roo_\Phi(\mu)$ обозначим характеристический многочлен $A_\Phi$, сокращенный на $\mu^4$. Тогда $\roo_\Phi(\mu)=\mu^2-\ld$, а знак $\ld$ определяет тип критической точки из $\mcm_\Phi$ по отношению к малой площадке, трансверсальной к $\mcm_\Phi \cap \mPel$ в $\mPel$. Для точки ранга 1 это полный тип, а точки ранга 0 это называют внешним типом по отношению к подсистеме $\mcm_\Phi$. Поскольку невырожденные точки ранга 0 принадлежат трансверсальному пересечению двух критических подсистем или трансверсальному самопересечению одной системы, то они получают два внешних типа, которые и определяют их полный тип в соответствующей приведенной системе.

Следующие утверждения относительно критических подсистем и множества критических точек отображения момента случай Ковалевской\,--\,Соколова получаются из результатов работ \cite{Ry2013,KhRCD2014} предельным переходом при обнулении второго поля, но, конечно, могут быть проверены и непосредственным вычислением. Для каждой из подсистем вводятся частные интегралы, обозначенные прописными буквами. Их константы обозначаются соответствующими строчными буквами (в том числе, с индексами). Для краткости индуцированной симплектической структурой мы называем 2-форму, индуцированную на четномерных подмногообразиях в $\mPel$ исходной симплектической структурой на $T SO(3)$ или, что то же самое, симплектической структурой на симплектических листах $\mPel$ скобки Пуассона. Мы явно указываем, где эта форма имеет точки вырождения. Более того, инвариантные соотношения выбраны в соответствии с замечанием~\ref{rem1} так, чтобы их скобка Пуассона была частным интегралом критической подсистемы, поэтому и множество точек вырождения описано в терминах таких интегралов.

\begin{propos}\label{propm1}
Множество $\mcm_1$, заданное системой
\begin{equation}\notag
Z_1^{(1)}=0,\quad Z_2^{(1)}=0,
\end{equation}
где
\begin{equation}\notag
\begin{array}{l}
\ds{Z_1^{(1)}=\frac{1}{4}(M_1^2-M_2^2)+\eps_1(\va_2M_3-\va_3M_2)+\eps_0\va_1-\eps_1^2a^2,}\\[5mm]
\ds{Z_1^{(2)}=\frac{1}{2}M_1M_2+\eps_1(\va_3M_1-\va_1M_3)+\eps_0\va_2},
\end{array}
\end{equation}
является критической подсистемой случая Ковалевской\,--\,Соколова, порожденной функцией $\Phi_1(h,\ell,k)=k$. На $\mcm_1$ определен частный интеграл $F_1=\{Z_1^{(1)},Z_2^{(1)}\}$:
\begin{equation}\notag
F_1=\frac{1}{4}(M_1^2+M_2^2)M_3+\eps_1(\va_1 M_2-\va_2 M_1)M_1+\eps_0\va_3 M_1-\eps_1^2 a^2 M_3,
\end{equation}
причем на $\mcm_1$ постоянные интегралов $F_1,H,L$ связаны соотношением
\begin{equation*}
  f_1^2=[\eps_0^2+2\eps_1^2(h+\eps_1^2a^2)][a^2(h+\eps_1^2a^2)-2\ell^2].
\end{equation*}
Нулевой уровень $F_1$ определяет множество точек вырождения индуцированной симплектической структуры в каждом $\mcm_1\cap \mPel$. Характеристический многочлен имеет вид
\begin{equation}\notag
\roo_{\Phi_1}(\mu)=\mu^2+4 f_1^2,
\end{equation}
в частности, при $f_1\ne 0$ все точки ранга $1$ из $\mcm_1$ невырождены и имеют тип ``центр''. Невырожденные точки ранга $0$, лежащие в $\mcm_1$, имеют внешний тип ``центр''.
\end{propos}

\begin{propos}\label{propm2}
Множество $\mcm_2$, заданное системой
\begin{equation}\notag
Z_1^{(2)}=0,\quad Z_2^{(2)}=0,
\end{equation}
где
\begin{equation}\notag
\begin{array}{l}
\ds Z_1^{(2)}=\frac{\sqrt{\eps_0^2+\eps_1^2(M_1^2+M_3^2)}}{\eps_1\sqrt{\eps_0^2+\eps_1^2M_3^2}}[\eps_1^2(M_3-2\eps_1\va_2)M_3-2\eps_0\eps_1^2\va_1+\eps_0^2],\\[5mm]
\ds Z_2^{(2)}=\frac{1}{\sqrt{\eps_0^2+\eps_1^2M_3^2}}[\eps_1^2(M_2+2\eps_1\va_2)M_3^2+2\eps_1^3\va_1M_1M_3+\\[5mm]
\qquad +\eps_0(\eps_0M_2-2\eps_1^2\va_2M_1+2\eps_0\eps_1\va_3)],
\end{array}
\end{equation}
является критической подсистемой случая Ковалевской\,--\,Соколова, порожденной функцией
\begin{equation}\notag
\Phi_2(h,\ell,k)=k-\frac{1}{4\eps_1^4}\left[2\eps_1^2(h+\eps_1^2 a^2)+\eps_0^2 \right]^2.
\end{equation}
На $\mcm_2$ определен частный интеграл $F_2=\eps_0 \{Z_2^{(1)},Z_2^{(2)}\}$:
\begin{equation}\notag
F_2=M_2 \sqrt{\eps_0^2+\eps_1^2(M_1^2+M_3^2)}[\eps_0^2+\eps_1^2(M_3-2\eps_1\va_2)^2].
\end{equation}
Нулевой уровень $F_2$ определяет в каждом $\mcm_2\cap \mPel$ точки вырождения индуцированной симплектической структуры. Характеристический многочлен имеет вид
\begin{equation}\notag
\roo_{\Phi_2}(\mu)=\mu^2+\frac{4}{\eps_0^2}f_2^2,
\end{equation}
в частности, при $f_2\ne 0$ все точки ранга $1$ из $\mcm_2$ невырождены и имеют тип ``центр''. Невырожденные точки ранга $0$, лежащие в $\mcm_2$, имеют внешний тип ``центр''.
\end{propos}

\begin{remark}\label{rem2}
{\small
Простое выражение для скобки $\{Z_2^{(1)},Z_2^{(2)}\}$ получено в точках $\mcm_2$ путем исключения переменных $\va_1,\va_3$ в силу инвариантных соотношений. В этом представлении легко видеть, что на $\mcm_2$ условие $F_2=0$ равносильно тому, что $M_2=0$. Таким же путем проверяется, что на $\mcm_2$ частным интегралом является функция
\begin{equation}\notag
Q=\frac{M_2}{\sqrt{\eps_0^2+\eps_1^2(M_1^2+M_3^2)}},
\end{equation}
и, кроме уравнения поверхности $\Pi_2$, не содержащего $\ell$, имеется следующее соотношение на постоянные первых интегралов
\begin{equation*}
8\eps_1^6 \ell^2-\eps_0^2 (\eps_0^2+2\eps_1^2 h)=\frac{\eps_1^2}{2} f_2 q.
\end{equation*}
}
\end{remark}

Следующую критическую подсистему удобно записать в переменных \eqref{eq6_5}. Обозначим
\begin{equation}\notag
\xi_1=Z_1^{(1)}+\ri Z_2^{(1)},\quad \xi_2=Z_1^{(1)}-\ri Z_2^{(1)}.
\end{equation}
В частности, интеграл $K$ примет вид $K=\xi_1\xi_2$.

\begin{propos}\label{propm3}
Замыкание $\mcm_3$ множества, заданного системой
\begin{equation}\notag
Z_1^{(3)}=0,\quad Z_2^{(3)}=0,
\end{equation}
где
\begin{equation}\notag
\begin{array}{l}
\ds Z_1^{(3)}=\sqrt{x_1 x_2}w_3-\frac{1}{\sqrt{x_1x_2}}\left[(w_1x_2+w_2x_1)z-\ri a^2\eps_1(x_1-x_2)\right],\\[5mm]
\ds Z_2^{(3)}=\frac{1}{\ri}\left(\frac{x_2}{x_1}\xi_1-\frac{x_1}{x_2}\xi_2\right),
\end{array}
\end{equation}
является критической подсистемой случая Ковалевской\,--\,Соколова, порожденной функцией
\begin{equation}\notag
\Phi_3(h,\ell,k)=k-\frac{[a^2(h+\eps_1^2 a^2)-2\ell^2]^2}{a^4}.
\end{equation}
На $\mcm_3$ определен частный интеграл
\begin{equation}\notag
F_3=\frac{1}{\sqrt{x_1x_2}}\left[(w_1+\ri\eps_1z)(w_2-\ri\eps_1 z)+\frac{1}{2}\left(\frac{x_2}{x_1}\xi_1+\frac{x_1}{x_2}\xi_2\right)\right]+\eps_1^2\sqrt{x_1x_2}
\end{equation}
такой, что
\begin{equation}\notag
\{Z_1^{(3)}, Z_2^{(3)}\}=2a^2 F_3.
\end{equation}
Нулевой уровень $F_3$ определяет в каждом $\mcm_3\cap \mPel$ точки вырождения индуцированной симплектической структуры.
\end{propos}

В этом случае вырожденные критические точки появляются и не в связи с вырождением индуцированной симплектической структуры.
\begin{propos}\label{propm31}
На инвариантном подмножестве $\mcm_3$ определен частный первый интеграл
\begin{equation}\notag
M =\frac{1}{2a^2}\left(\frac{x_2}{x_1}\xi_1+\frac{x_1}{x_2}\xi_2\right),
\end{equation}
такой, что постоянные общих интегралов связаны с постоянными интегралов $F_3,M$ соотношениями
\begin{equation}\notag
\begin{array}{c}
\ds  \ell^2=\frac{a^2}{4(m+\eps_1^2)}(f_3^2-\eps_0^2), \qquad k=a^4 m^2, \\[3mm]
\ds  h=-a^2(m+\eps_1^2)+\frac{1}{2(m+\eps_1^2)} (f_3^2-\eps_0^2)
\end{array}
\end{equation}
и характеристический многочлен имеет вид
\begin{equation}\notag
\roo_{\Phi_3}(\mu)=\mu^2 + 4 a^4 m f_3^2.
\end{equation}
В частности, при $f_3\ne 0$ и $m \ne 0$ все точки ранга $1$ из $\mcm_3$ невырождены и имеют тип ``центр'' при $m>0$ и тип ``седло'' при $m<0$. Для невырожденных точек ранга $0$ эти же условия определяют внешний тип по отношению к $\mcm_3$.
\end{propos}


Устремляя $\eps_1$ к нулю, нетрудно убедиться, что критическая подсистема $\mcm_1$ есть аналог первого класса Аппельрота случая Ковалевской, критическая подсистема $\mcm_3$ есть аналог второго и третьего классов. Критическая подсистема $\mcm_2$ является новой, она не имеет такого аналога и непуста лишь при условии $\eps_1^2 > \zeta_1$. Так называемые особо замечательные движения четвертого класса Аппельрота представляют собой наиболее устойчивую подсистему относительно различных обобщений случая Ковалевской. К ним относятся, например, случай Бобылева\,--\,Стеклова и его обобщение на гиростат весьма общего вида \cite{PVLect}, обобщения на твердое тело в двойное поле \cite{Yeh2,Kh2006} и на гиростат в двойном поле \cite{KhND07}. Аналог этого класса движений есть и в рассматриваемой задаче. Порождающую функцию $\Phi_4(h,\ell,k)$ можно получить, исключая из уравнений поверхности $\Pi_4$ параметр $s$, однако, в таком виде она, как и в классическом случае Ковалевской, оказывается бесполезной.

\begin{propos}\label{propm4}
Множество $\mcm_4$, заданное системой
\begin{equation}\notag
Z_1^{(4)}=0,\quad Z_2^{(4)}=0,
\end{equation}
где
\begin{equation}\notag
\begin{array}{l}
\ds Z_1^{(4)}=M_2, \quad \ds  Z_2^{(4)}=\frac{1}{2}M_1 M_3  -\eps_1 \va_2 M_1+\eps_0 \va_3,
\end{array}
\end{equation}
является критической подсистемой случая Ковалевской\,--\,Соколова.
На $\mcm_4$ имеется частный интеграл
\begin{equation}\notag
S=-\frac{2\eps_0 L}{M_1},
\end{equation}
такой, что постоянные общих интегралов и интеграла $S$ связаны уравнениями поверхности $\Pi_4$ и
\begin{equation}\notag
\{Z_1^{(4)}, Z_2^{(4)}\}=2 H -3 S.
\end{equation}
Условие $2H-3S=0$ определяет точки вырождения индуцированной симплектической структуры в каждом $\mcm_4\cap \mPel$. Геометрически оно задает ребро возврата поверхности $\Pi_4$.
\end{propos}

Для вычисления типов критических точек порождающая функция $\Phi_4$ слишком сложна. Для нас важно следующее из уравнений поверхности $\Pi_4$ соотношение на дифференциалы общих интегралов
\begin{equation}\notag
  \frac{4 (\eps_0^2+2\eps_1^2 s)\ell}{s} dL+dK-2(h+\eps_1^2a^2-s) dH=0.
\end{equation}
Учитывая, что $L$ есть функция Казимира скобки Пуассона и, значит, $\sgrad L \equiv 0$, рассмотрим функцию
\begin{equation}\notag
  \tilde \Phi_4 = k - 2(h+\eps_1^2a^2-s) h.
\end{equation}
Очевидно, что на $\mcm_4$ тождественно $\sgrad \tilde \Phi_4 =0$, где косой градиент берется, конечно, от композиции $\tilde \Phi_4$ с отображением момента. Более того, для упрощения процедуры следует до вычисления косого градиента и его линеаризации (то есть искомого симплектического оператора) считать коэффициент $2(h+\eps_1^2a^2-s)$ произвольной константой, и лишь после вычисления подставить в него функции $H,S$.

\begin{propos}\label{propm41}
На инвариантном подмножестве $\mcm_4$ характеристический многочлен симплектического оператора имеет вид
\begin{equation*}
\roo_{\tilde \Phi_4}(\mu)=\mu^2 - \ld_4,
\end{equation*}
где
\begin{equation}\notag
\ld_4 = 2(\eps_0^2+2\eps_1^2s)(3s-2h)(\eps_0^2a^2-2h s+2s^2).
\end{equation}
В частности, при $\ld_4 \ne 0$ все точки ранга $1$ из $\mcm_4$ невырождены и имеют тип ``центр'' при $\ld_4 <0$ и тип ``седло'' при $\ld_4>0$. Для невырожденных точек ранга $0$ эти же условия определяют внешний тип по отношению к $\mcm_4$.
\end{propos}

\begin{theorem}\label{theo6}
Множество $\mathcal{C}$ критических точек отображения момента случая Ковалевской\,--\,Соколова является объединением инвариантных подмножеств, служащих фазовыми пространствами перечисленных выше четырех критических подсистем,
\begin{equation*}
  \mathcal{C}=\bigcup_{i=1}^4 \mcm_i.
\end{equation*}
Невырожденные точки ранга $0$ являются трансверсальными пересечениями двух открытых подмножеств критических подсистем и тип точек ранга $0$ определяются парой внешних типов, вычисленных для критических подсистем.
\end{theorem}

Для доказательства отметим вначале, что критические точки произвольной функции $f$ на $\mP^5$ определяются уравнениями (см. лемму~\ref{lem4})
\begin{equation}\label{eq7_3}
  \partial _\bm f =0, \qquad  {\bf D} f =0.
\end{equation}
Критические точки $H$ уже изучены. Поэтому в линейных комбинациях дифференциалов можно один из коэффициентов при остальных интегралах считать ненулевым. Записывая условия \eqref{eq7_3} для комбинации вида $K-\sigma H$, получим подсистемы $\mcm_1,\mcm_2$. Таким образом, эти подсистемы отражают зависимость $K$ и $H$ без учета их ограничения на уровень функции $L$. В частности, для оставшихся критических точек мы можем считать ненулевым коэффициент при дифференциале $L$. Вводя функцию $f=L-\tau H -\sigma K$, запишем условия \eqref{eq7_3} и исключим из них $\tau, \sigma$. Получим систему инвариантных соотношений, ранг которой в искомых точках должен равняться двум. Легко проверить, что каждое из этих уравнений можно записать в двух видах
\begin{equation*}
  Z_1^{(i)} f_{1i} + Z_2^{(i)} f_{2i} =0  \quad (i=3,4).
\end{equation*}
Вначале запишем систему для $i=3$, затем, предполагая, что вектор  $(Z_1^{(3)}, Z_2^{(3)})$ ненулевой, исключим эти функции. Полученные уравнения представим в том же виде для $i=4$. Предполагая вектор $(Z_1^{(4)}, Z_2^{(4)})$ ненулевым, исключим и эти функции. Возьмем любое независимое уравнение вида $F=0$ из оставшихся и убедимся, что в силу исходной системы дифференциальных уравнений $\ddot F$ не обращается в нуль тождественно при $F=0$ и $\dot F=0$. Таким образом, ранг оставшейся системы инвариантных соотношений больше двух, и она может порождать только критические точки ранга 0, которые уже известны.

Трансверсальность пересечений (или самопересечений) критических подсистем в невырожденных точках ранга 0 следует из существования двух вычисленных выше внешних типов. Теорема доказана.

\section{Примеры изоэнергетических диаграмм и грубая топология}\label{sec9}

В силу того, что уровни энергии компактны, что следует, например, из представления функции $H$ в виде \eqref{eq2_1}, удобно рассматривать бифуркационные диаграммы ограничения отображения момента на изоэнергетические уровни, то есть ограничения отображения $L{\times}K$ на четырехмерные многообразия $\mP^5 \cap \{H=h\}$. Конечно, такие диаграммы -- это сечения общей бифуркационной диаграммы отображения момента плоскостями $h={\rm const}$. С этой целью и уравнения несущих поверхностей $\Pi_1 - \Pi_4$ записаны так, чтобы в них величина $h$ играла роль параметра. Для описания грубой топологии мы приводим оснащенные диаграммы, то есть такие, в которых на каждом ребре указан атом происходящей бифуркации, а в непустых клетках дополнения указано количество торов Лиувилля.

\begin{figure}[!ht]
\includegraphics[width=75mm,keepaspectratio]{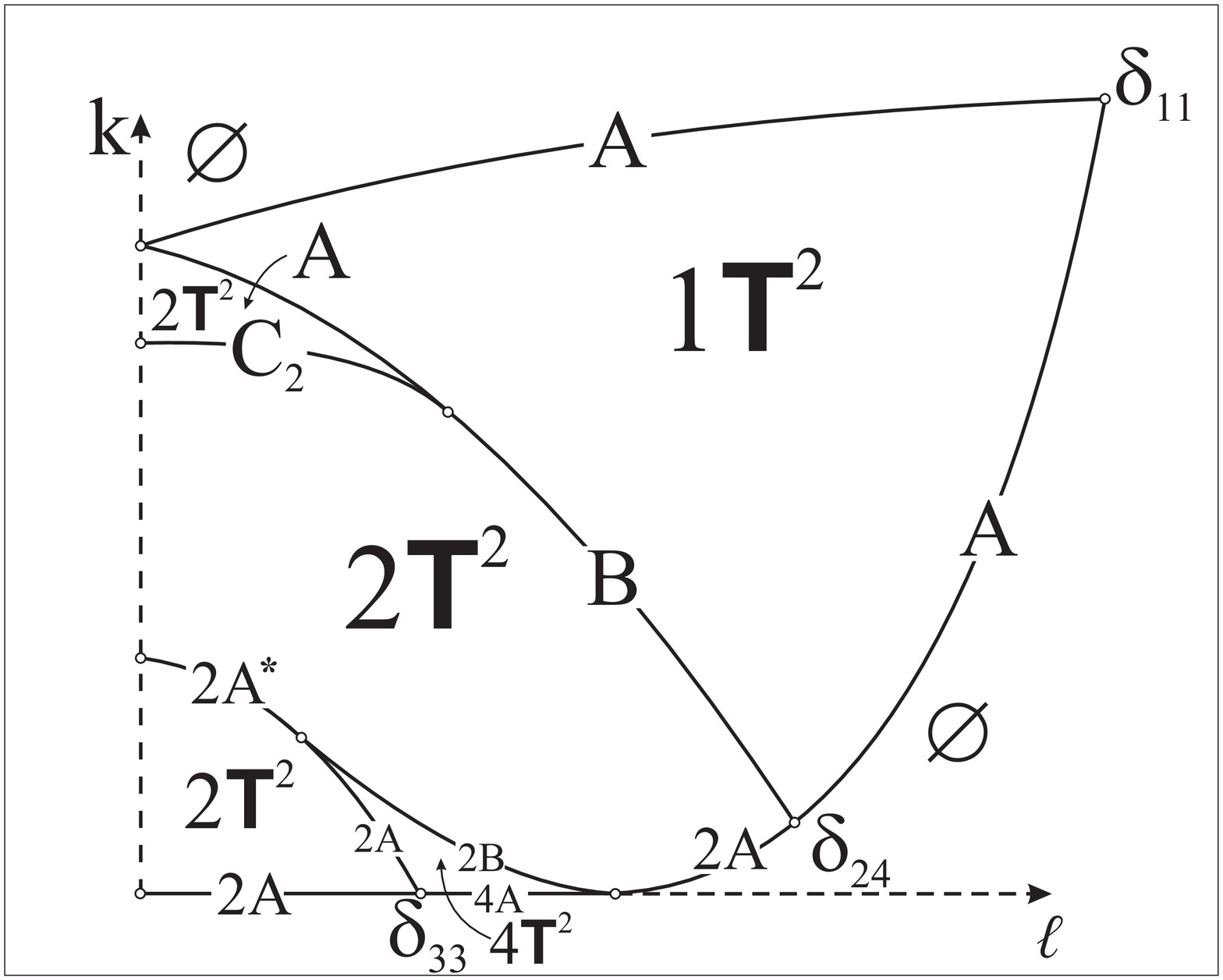} \hspace{5mm} \includegraphics[width=75mm,keepaspectratio]{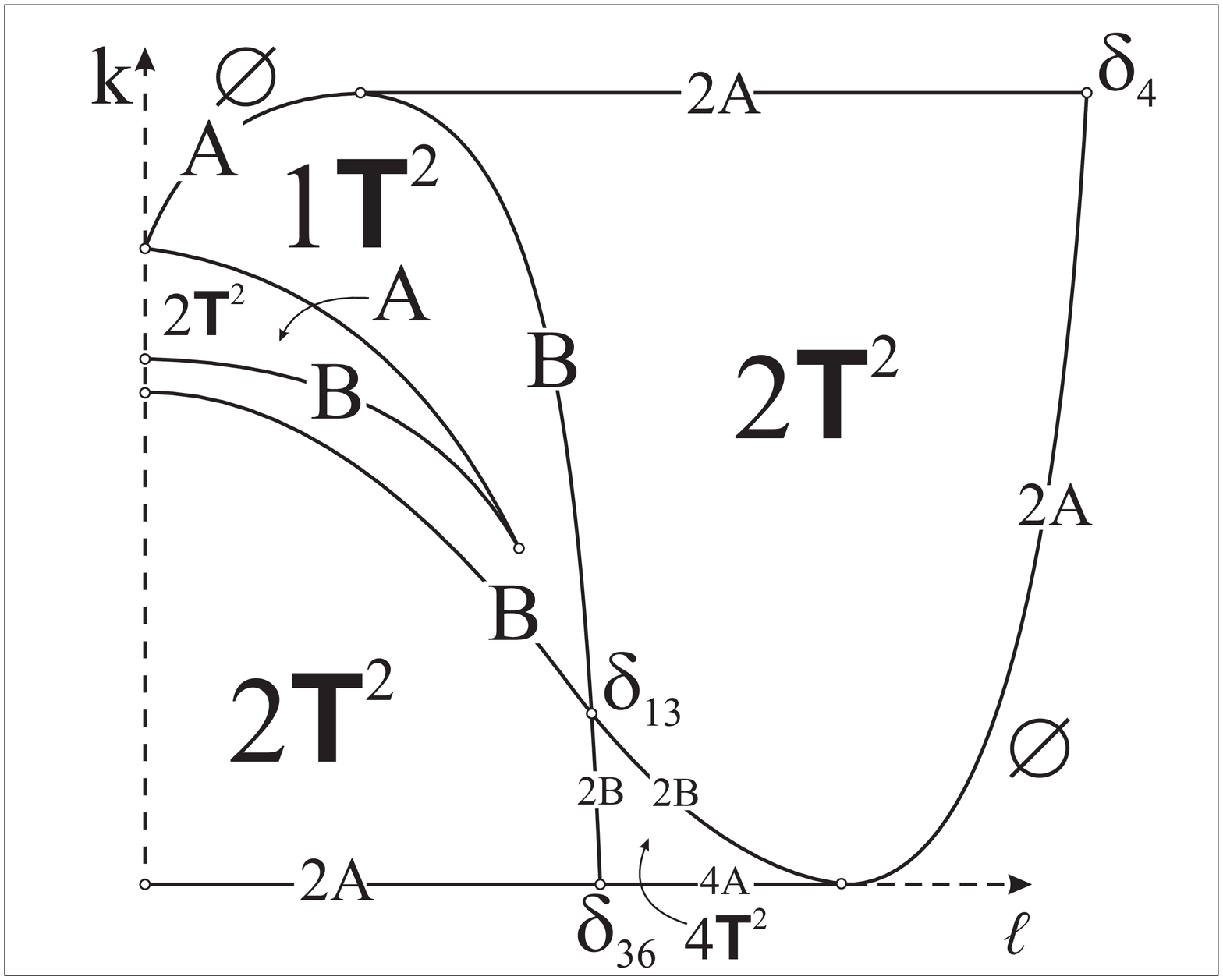}\\
\includegraphics[width=75mm,keepaspectratio]{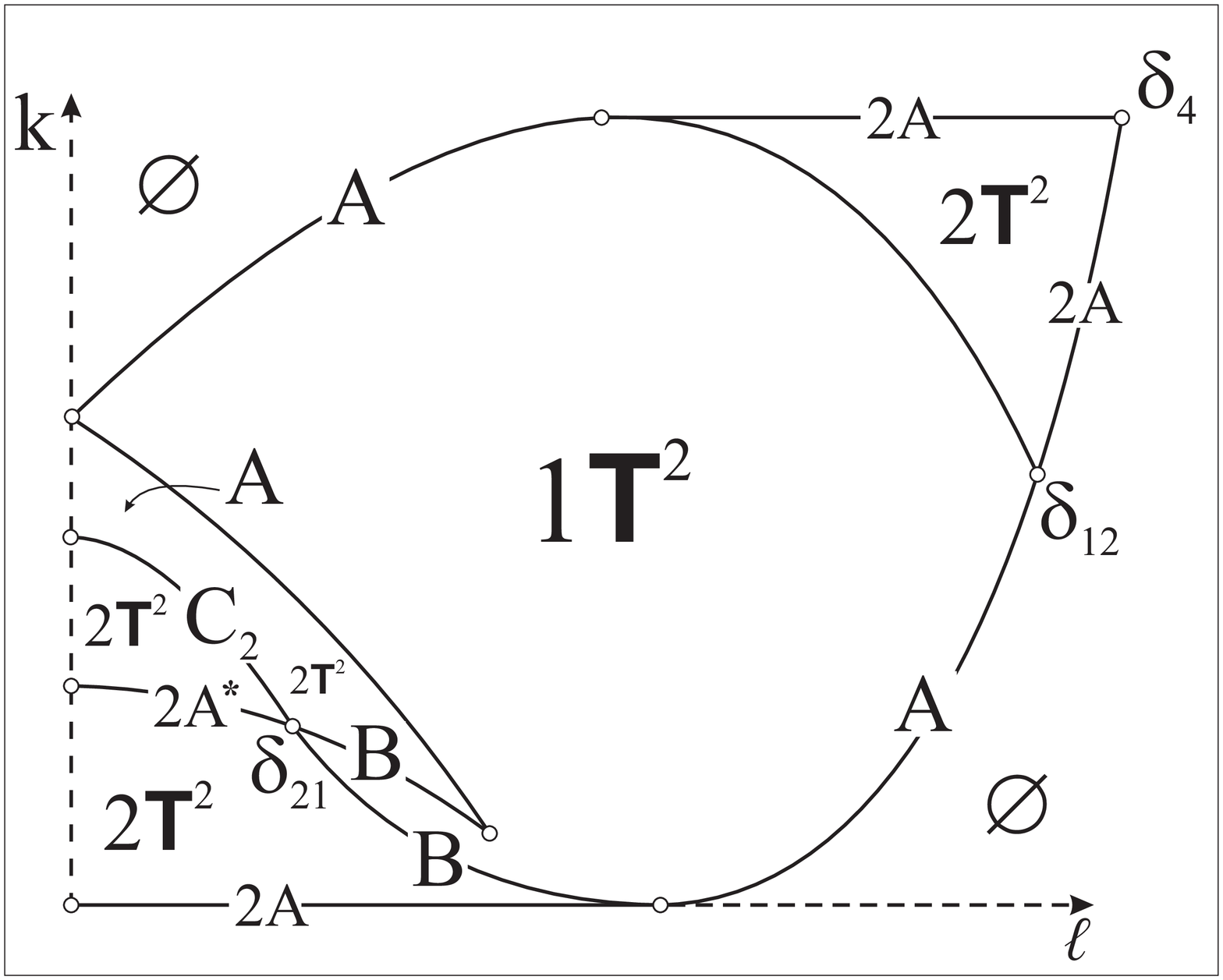} \hspace{5mm} \includegraphics[width=75mm,keepaspectratio]{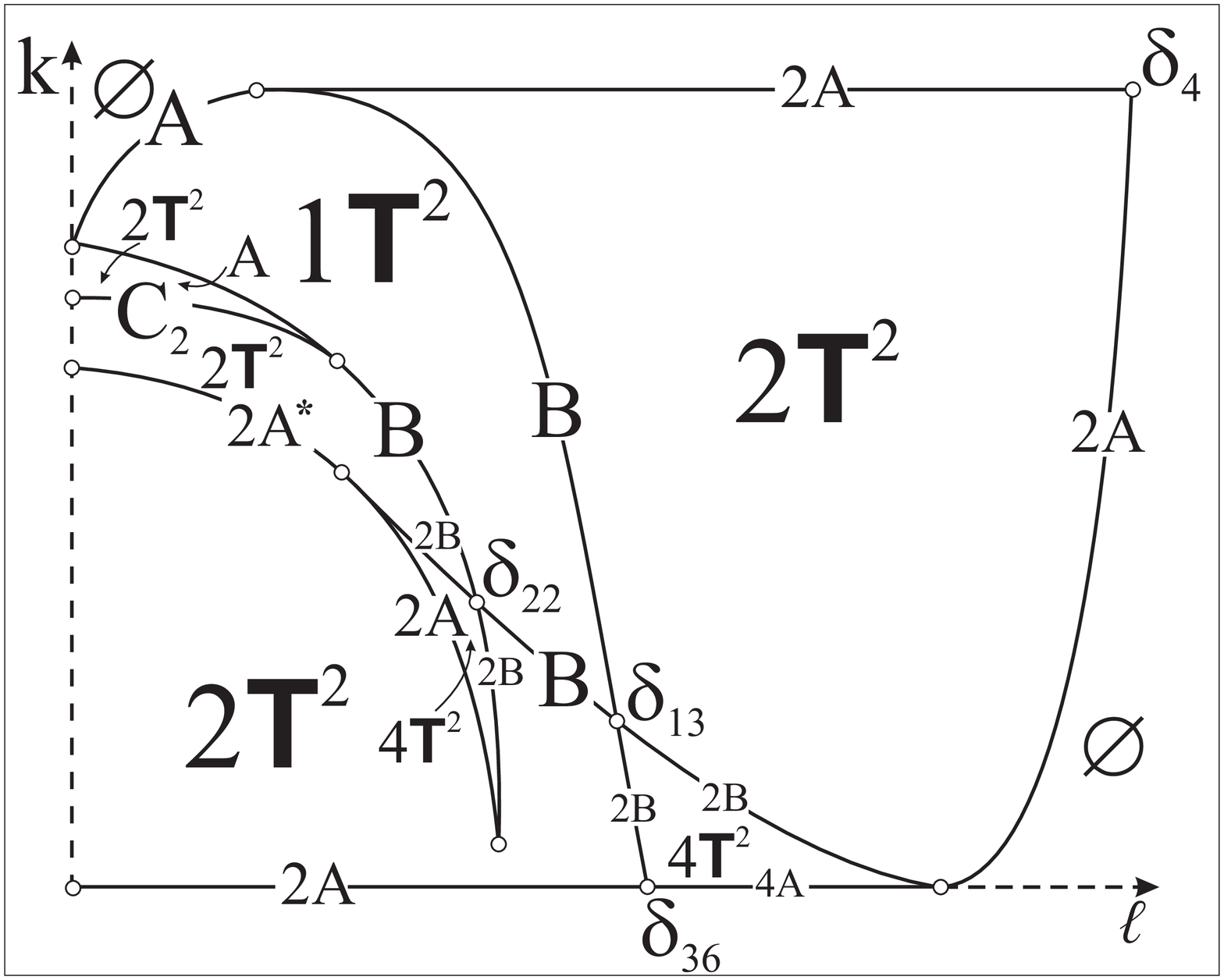}
\caption{Примеры оснащенных бифуркационных диаграмм}
\label{figbifs1}
\end{figure}

\begin{figure}[!ht]
\includegraphics[width=75mm,keepaspectratio]{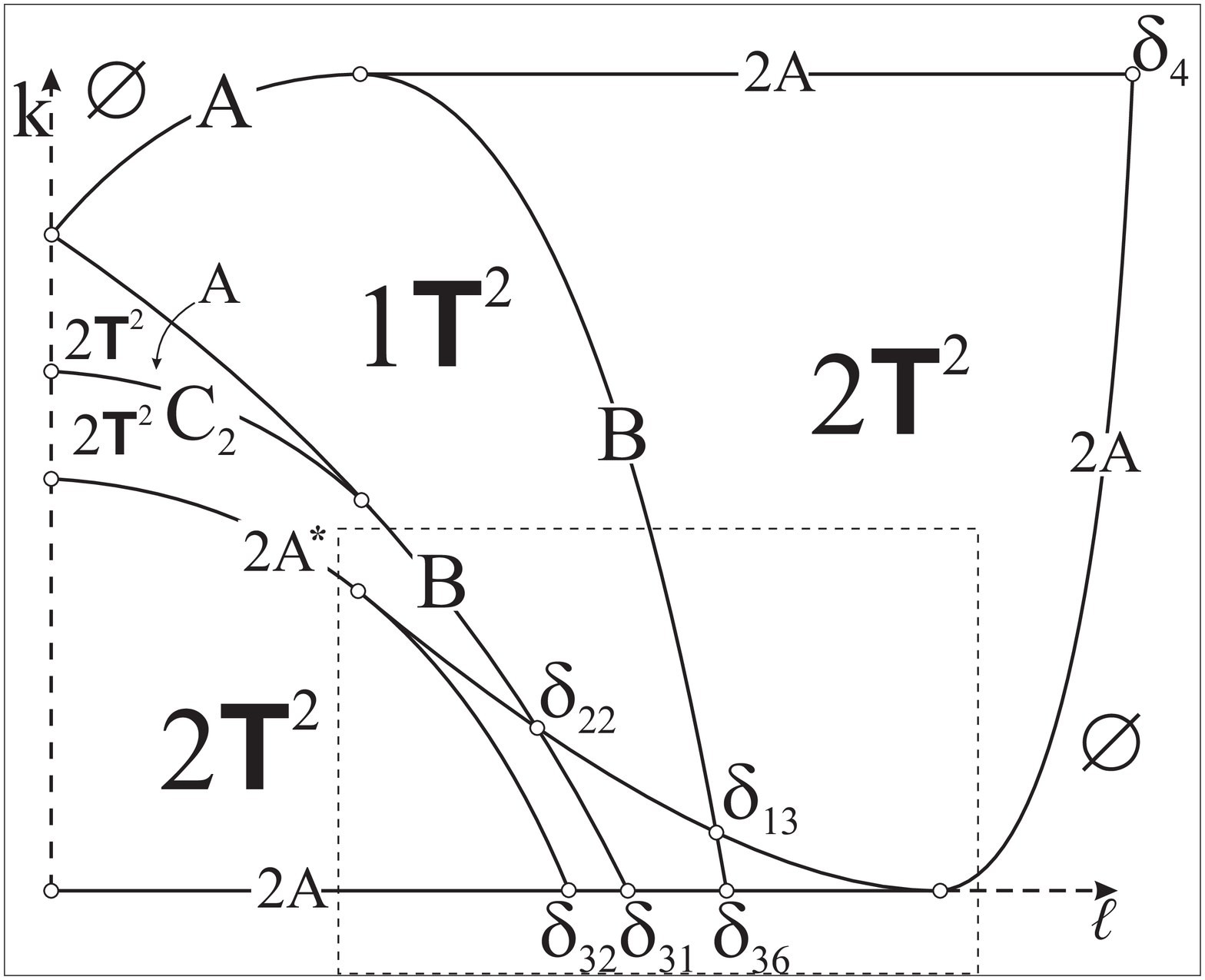} \hspace{5mm} \includegraphics[width=75mm,keepaspectratio]{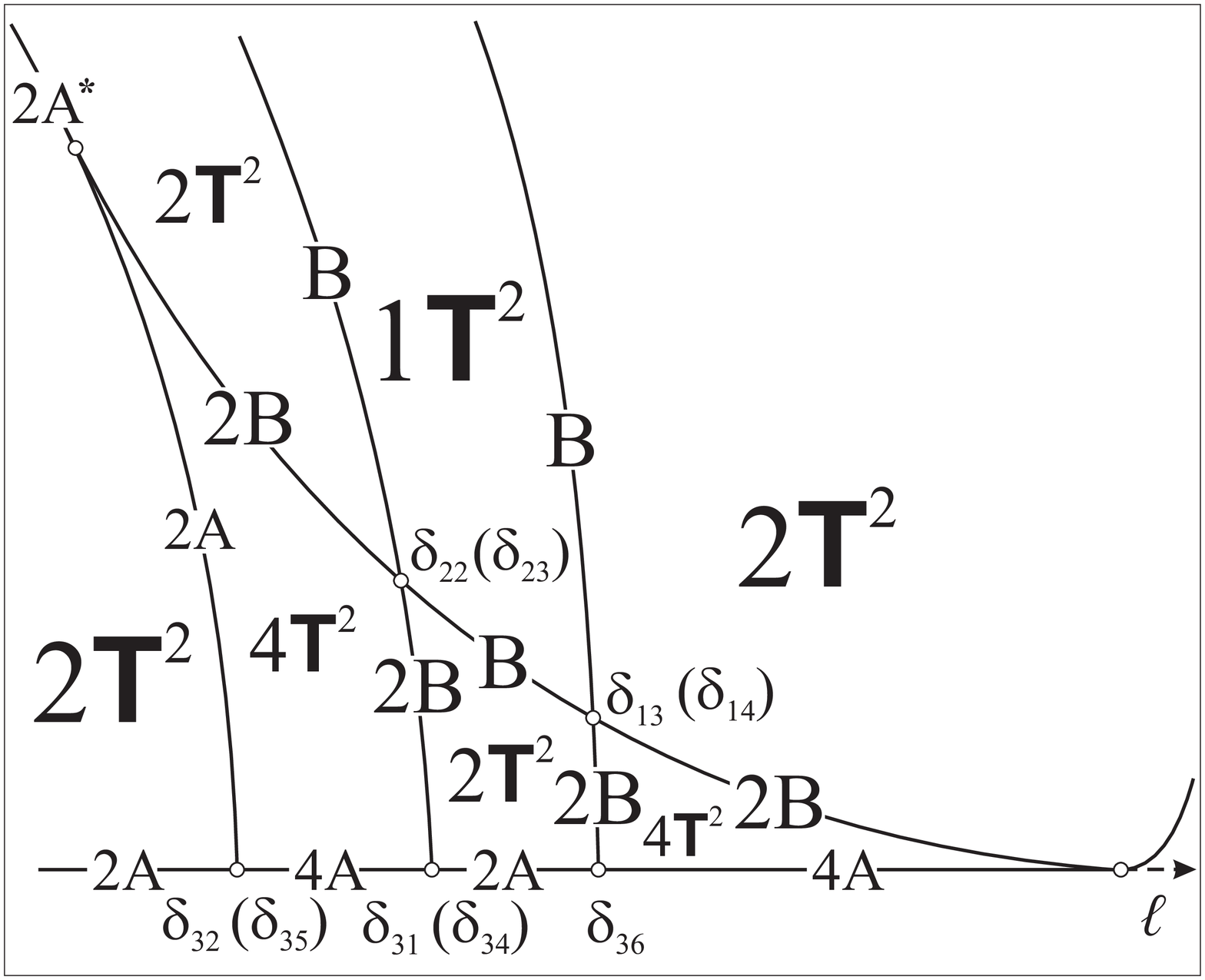}
\caption{Пример оснащенной бифуркационной диаграммы с фрагментом}
\label{figbifs3}
\end{figure}

Полная классификация выходит за рамки настоящей статьи, являясь технически довольно громоздкой задачей, хотя алгоритм ее решения известен. Считая, как и ранее, параметры $\eps_0, a$ заданными, рассмотрим кривые на плоскости $(\eps_1,h)$, которые служат образами вырожденных точек ранга 0 и экстремальных значений $H$ на семействах вырожденных решений ранга 1 \cite{KhRy2011}. Полученные кривые оказываются разделяющим множеством при классификации изоэнергетических диаграмм. Так, для волчка в двойном поле таких диаграмм получилось 19 (см. \cite{Kh34}), а для гиростата Ковалевской\,--\,Яхья имеется 33 диаграммы (см. \cite{mtt40,RyabHarlUdgu2}, все иллюстрации приведены в обзоре \cite{arxiv}). На рис.~\ref{figbifs1} -- \ref{figbifs3} приведены примеры оснащенных диаграмм, на которых видны все найденные выше невырожденные критические точки ранга 0. Естественно, диаграммы в целом симметричны относительно оси $Ok$. Типы бифуркаций в седловых особенностях ранга 1 определяются предельным переходом из задачи о гиростате в двойном поле \cite{Ry2013}.

Отметим, что представители пар классов $(\dd_{13},\dd_{14})$, $(\dd_{22},\dd_{23})$, $(\dd_{31},\dd_{34})$ и тройки классов $(\dd_{32},\dd_{33},\dd_{35})$ как точки ранга 0 на самом деле не отличаются: точку одного из таких классов можно непрерывно перевести в расширенном фазовом пространстве в точку другого класса, не пересекая множества вырожденных точек ранга 0. На рисунке фрагмента варианты расположения точек в эквивалентных по существу диаграммах указаны в скобках. В плоскости параметров $(\zu,\wu)$ такие классы разделены следами точек $\po_1,\po_2,\po_3$ трансверсального пересечения кривых $\dd_i$. На изоэнергетических диаграммах они различаются порядком проекций всего набора точек, отвечающих относительным равновесиям, на ось $O\ell$.  Глобальное топологическое отличие здесь в том, что при пересечении соответствующих значений $(\ell,h)$ происходят, как показано выше в табл.~1, различные бифуркации трехмерных изоэнергетических многообразий приведенных систем, а изменение порядка проекций влечет изменение набора тонких инвариантов Фоменко\,--\,Цишанга, отвечающих фиксированному значению энергии.

\begin{figure}[!ht]
\centering
\includegraphics[width=0.45\textwidth,keepaspectratio]{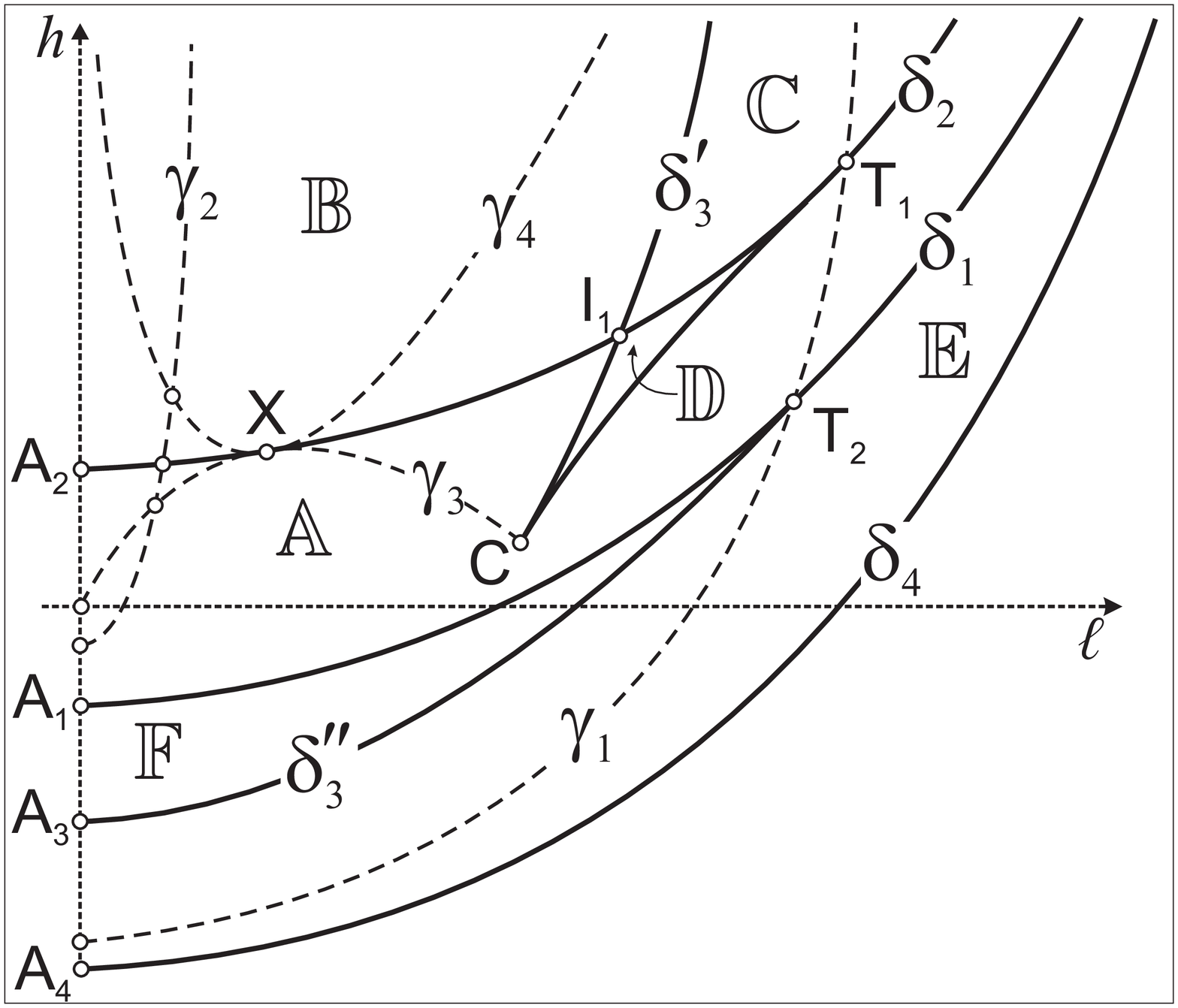}\hspace{5mm}
\includegraphics[width=0.45\textwidth,keepaspectratio]{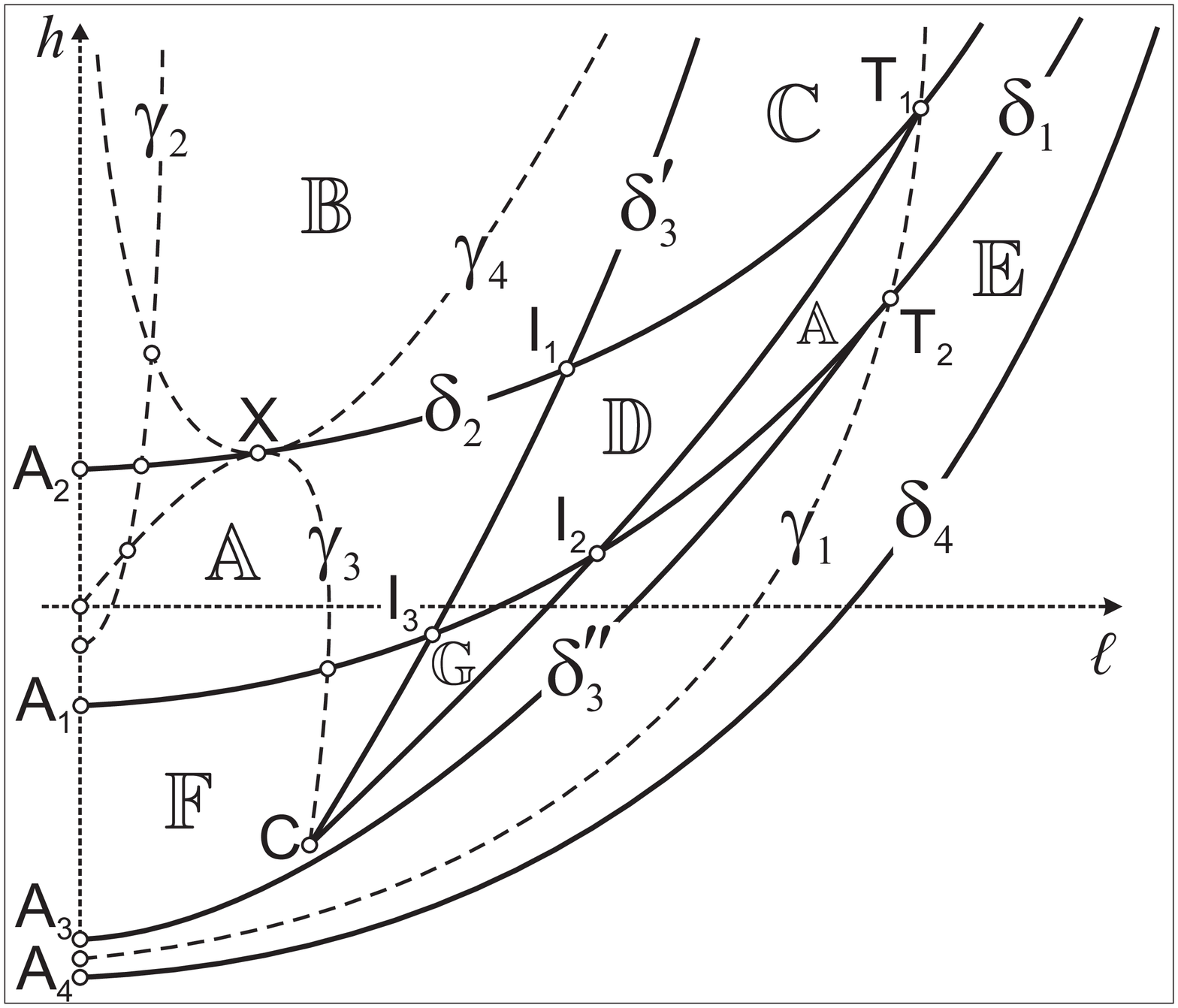}
\caption{Примеры диаграмм Смейла\,--\,Фоменко}\label{figsmfom}
\end{figure}

Рассматривая сечения изоэнергетических оснащенных диаграмм прямыми $\ell={\rm const}$, получим соответствующие грубые инварианты Фоменко на трехмерных изоэнергетических уровнях $\iso$ приведенных систем. Основные примеры таких инвариантов можно извлечь из представленных рисунков. Для их полной классификации нужно на плоскость $(\ell,h)$ нанести в дополнение к диаграммам Смейла кривые, которые являются образами вырожденных точек ранга 1. Из предложения \ref{propm1}, замечания~\ref{rem2} и предложений~\ref{propm31}, \ref{propm41} следуют уравнения этих кривых:
\begin{eqnarray}
  \gamma_1&:& 2\ell^2-a^2(h+\eps_1^2 a^2)=0, \nonumber\\
  \gamma_2&:& 8\eps_1^6 \ell^2-\eps_0^2 (\eps_0^2+2\eps_1^2 h)=0, \nonumber\\
  \gamma_3&:& 4 h^3 -27 \eps_0^2 \ell^2 =0, \nonumber\\
  \gamma_4&:& 8\ell^4-4 a^2 h \ell^2 +\eps_0^2 a^6 =0. \nonumber
\end{eqnarray}
Полученное множество называют диаграммой Смейла\,--\,Фоменко. Эта диаграмма служит разделяющим множеством для различных инвариантов Фоменко\,--\,Цишанга. Отметим, что теперь иначе объяснимо и появление точки $X$ на диаграммах Смейла -- в ней сходятся две последние разделяющие кривые. Примеры диаграмм Смейла\,--\,Фоменко показаны на рис.~\ref{figsmfom}.

Классификация всех диаграмм Смейла\,--\,Фоменко также возможна, но технически сложна и выходит за рамки настоящей работы. Пример такой классификации (для случая Ковалевской\,--\,Яхья) имеется в работе \cite{KhRy2011}.

Теперь, исходя из грубой топологии четырехмерных изоэнергетических уровней, представленной оснащенными изоэнергетическими диаграммами, устанавливаем вид 4-атома (и, соответственно, круговой молекулы) для всех точек ранга 0 типа ``седло-седло'', то есть в тех случаях, где такой 4-атом не определен однозначно типом самой точки. В данном случае, получаем атом $B{\times}B$ для точек $\dd_{13},\dd_{14},\dd_{22},\dd_{23}$ и атом $(B{\times}C_2)/\mathbb{Z}_2$ для точек единственного класса $\dd_{21}$.

В заключение отметим, что все полученные здесь результаты, в силу эквивалентности задач, автоматически переносятся на задачу о волчке Ковалевской на коалгебре $so(3,1)^*$. В описании критического множества отображения момента имеется определенное соответствие с уравнениями для волчка Ковалевской на $so(4)^*$, полученными в работе \cite{KozIK}, однако, последние не представлены в ``неприводимой'' форме систем, количество уравнений в которых совпадает с рангом, и не позволяют явным образом вычислять типы критических точек.

\section{Благодарности} Авторы благодарят профессора М.П. Харламова за постоянное внимание к работе и ценные советы, касающиеся как содержания, так и методологии исследования.

\end{document}